\documentclass[twocolumn]{aastex631}

\shorttitle{Methanol mapping in cold cores}
\shortauthors{Punanova et al.}
\graphicspath{{./}{figures/}}

\begin{document}

\title{Methanol Mapping in Cold Cores: Testing Model Predictions\footnote{This work is based on observations carried out under projects 013-18, 125-18 and 031-19 with the IRAM~30~m telescope. Institut de Radioastronomie Millim\'etrique (IRAM) is supported by INSU/CNRS (France), MPG (Germany) and IGN (Spain).}}


\correspondingauthor{Anna Punanova}
\email{anna.punanova@urfu.ru, punanovaanna@gmail.com}

\author[0000-0001-6004-875X]{Anna Punanova}
\affiliation{Ural Federal University, 620002, 19 Mira street, Yekaterinburg, Russia}

\author[0000-0003-1684-3355]{Anton Vasyunin}
\affiliation{Ural Federal University, 620002, 19 Mira street, Yekaterinburg, Russia}

\author[0000-0003-1481-7911]{Paola Caselli}
\affiliation{Max-Planck-Institut f\"ur extraterrestrische Physik, Giessenbachstrasse 1, 85748 Garching, Germany}

\author[0000-0002-3900-1313]{Alexander Howard}
\affiliation{School of Physics and Astronomy, Cardiff University, 5~The Parade, Cardiff, CF24 3AA, UK}

\author[0000-0002-6787-5245]{Silvia Spezzano}
\affiliation{Max-Planck-Institut f\"ur extraterrestrische Physik, Giessenbachstrasse 1, 85748 Garching, Germany}

\author{Yancy Shirley}
\affiliation{Steward Observatory, University of Arizona, Tucson, AZ~85721, USA}

\author[0000-0002-9485-4394]{Samantha Scibelli}
\affiliation{Steward Observatory, University of Arizona, Tucson, AZ~85721, USA}

\author[0000-0002-1189-9790]{Jorma Harju}
\affiliation{Department of Physics, P.O. BOX 64, FI-00014 University of Helsinki, Finland}



\begin{abstract}

Chemical models predict that in cold cores gas-phase methanol is expected to be abundant at the outer edge of the CO depletion zone, where CO is actively adsorbed. CO adsorption correlates with volume density in cold cores, and, in nearby molecular clouds, the catastrophic CO freeze-out happens at volume densities above 10$^4$ cm$^{-3}$. The methanol production rate is maximized there and its freeze-out rate does not overcome its production rate, while the molecules are shielded from UV~destruction by gas and dust. Thus, in cold cores, methanol abundance should generally correlate with visual extinction that depends both on volume and column density. In this work, we test the most basic model prediction that maximum methanol abundance is associated with a {\it local} $A_V\sim4$~mag in dense cores and constrain the model parameters with the observational data. With the IRAM~30~m antenna, we mapped the CH$_3$OH (2--1) and (3--2)  transitions toward seven dense cores in the L1495 filament in Taurus to measure the methanol abundance. We use the {\it Herschel}/SPIRE maps to estimate visual extinction, and the C$^{18}$O(2--1) maps from \citet{Tafalla2015} to estimate CO depletion. We explored the observed and modeled correlations between the methanol abundances, CO depletion, and visual extinction varying the key model parameters. The modeling results show that hydrogen surface diffusion via tunneling is crucial to reproduce the observed methanol abundances, and the needed reactive desorption efficiency matches the one deduced from laboratory experiments.
\end{abstract}



\section{Introduction} \label{sec:intro}

Methanol is a key precursor for many complex organic and pre-biotic molecules found in regions of star- and planet formation and thus it is very important for the growth of molecular complexity in the interstellar medium. It is observed at all stages of star formation, including the dense cold molecular gas within starless cores, characterized by temperatures of $T\simeq10$~K, gas densities $n({\rm H_2})>10^4$~cm$^{-3}$, and subsonic turbulence \citep[e.g.,][]{Benson1989,Bergin2007,Keto2008}. \citet{Geppert_ea06} showed that gas-phase synthesis cannot account for the observed amounts of methanol in the cold gas, that implies that chemical processes on interstellar dust grains must play the major role in the formation of this molecule. Indeed, the laboratory studies by~\citet[][]{Watanabe2002} and \citet[][]{Fuchs_ea09} confirmed that methanol can be formed efficiently during the hydrogenation of CO molecules on surfaces of interstellar dust analogs at temperatures of $\sim$10~K. Once formed, part of the solid methanol is delivered from cold grains to the gas phase where it is widely observed. The delivery mechanism is not well understood. In cold dark star-forming clouds, methanol is most likely delivered to the gas phase via the so-called reactive desorption mechanism~\citep[][]{Garrod2007,Minissale2016,Chuang_ea18}. During reactive desorption events, a fraction of the energy released in exothermic surface reactions is spent by the formed molecule to overcome the Van der Waals force which bind it to the surface; in this way, reactive desorption occurs. The efficiency of reactive desorption, i.e. the probability that a reaction product will be released to the gas, depends on a number of factors including the exothermicity of a reaction, the properties of the underlying surface etc. Existing studies report very different values of the efficiency of reactive desorption for species and surfaces, including methanol and other products of CO hydrogenation~\citep[see e.g.,][]{Garrod2007,Vasyunin2013,Minissale2016,Fredon_ea17,Vasyunin2017,Wakelam_ea17}. Unlike water ice mantles, surfaces rich in CO and CH$_3$OH are needed to have efficient reactive desorption, as found by \citet{Minissale2016}. Observations of cold dense cores show strong depletion of CO molecule from the gas phase \citep[][]{Willacy_ea98,Caselli1999,Crapsi2005,Pagani2007}. CO molecules efficiently freeze-out on cold dust grains, thus making them efficient chemical reactors producing methanol and other products of hydrogenation of carbon monoxide. Given the dynamical quiescence and near-spherical geometry of pre-stellar cores, one can utilize them as natural laboratories to study the poorly known details of methanol formation and its link to carbon monoxide. 

Methanol emission toward cold dense cores is observed in integrated intensity maps as ring-like structures \citep[see, e. g.,][]{Tafalla2006,Bizzocchi2014,Punanova2018a,Harju2020,Spezzano_ea20} or a single peak toward the core center \citep{Nagy2019} if the cores have not experienced substantial CO freeze-out, maybe because of their relative dynamical youth compared to other dense cores. Methanol distribution within the ring-like structures is often inhomogeneous \citep[see e.g.,][]{Bizzocchi2014,Harju2020}. \citet{Jimenez-Serra2016} showed that abundances of methanol and other complex organic molecules (COMs) in the L1544 pre-stellar core are higher in the core shell, that corresponds to $A_V\sim$~7.5--8~mag \citep[see also][]{Vastel2014}. The model, presented in \citet{Vasyunin2017}, predicts that the maximal gas-phase abundance of the organic species is at $A_V\sim$~8~mag. \citet{Scibelli2020} presented a methanol survey toward the L1495 filament and detected methanol down to the line-of-sight $A_V=3$~mag, with 70$^{\prime\prime}$ beam of the Arizona Radio Observatory ARO~12~m antenna. In this work, we first test the most basic model prediction that maximum methanol abundance is associated with $A_V\simeq8$~mag on the line of sight in dense cores (i.e. a local $A_V$ within the core of $\sim$4~mag). Second, we attempt to put observational constraints on the parametrizations of reactive desorption used in chemical models of pre-stellar cores. The advantage of this study is the wealth of spatial data from high-resolution mapping of seven dense cores embedded within the same filament L1495. This allowed us to constrain model parameters on a multitude of data points, in contrast to a number of previous studies.

The targets of our study are the cold cores in the well studied filamentary structure L1495 in the Taurus molecular cloud \citep{Lynds1962}, which is a nearby \citep[130--135 pc distant;][]{Schlafly2014,Roccatagliata2020}, quiescent low-mass star-forming region. The filament contains tens of dense cores \citep{Marsh2014}, and about fifty low-mass protostars in different evolutionary stages \citep{Rebull2010}. \citet{Hacar2013} and \citet{Seo2015} note that some parts of the filament are young (B211 and B216) and others (B213 and B7) are more evolved and actively star-forming, based on the amount of embedded protostars, their class and the level of gas turbulence. The dense cores embedded in the filament were detected in N$_2$H$^+$ \citep{Hacar2013}, dust continuum emission \citep{Marsh2014} and in ammonia \citep{Seo2015}. The dense cores show low gas temperature decreasing toward their centers \citep[8--12~K;][]{Seo2015}, subthermal gas motions \citep{Seo2015,Punanova2018b}, coherent velocity structure, and slow rotation \citep{Punanova2018b}. The recent observations of the methanol lines at 96.7~GHz toward 31~starless cores detected by \citet{Seo2015} show weaker detections of methanol toward the more evolved regions, and the highest methanol gas abundance in the outskirts of dense gas in B211 \citep{Scibelli2020}. 

This paper presents maps of methanol lines at 2 and 3~mm toward seven dense cores in the L1495 filament to study the chemical connection between gas phase methanol, visual extinction and CO depletion in molecular clouds. We use the observational results to constrain free parameters of the chemical model by \citet{Vasyunin2017} and test the model predictions. In Section~\ref{sec:obs}, we present the details of observations and data reduction. In Section~\ref{sec:results}, we present the results of methanol column density measurements, abundance estimations, CO depletion, molecular hydrogen and visual extinction estimations, and describe the chemical modeling. In Section~\ref{sec:discussion} we discuss the results and correlations between methanol abundance and visual extinction and CO depletion, comparison with the chemical model results and constraining of the model parameters. The conclusions are given in Section~\ref{sec:conclusions}.

\section{Observations and Data Reduction} \label{sec:obs}

\begin{table}
\caption{The mapped cores. The numbers are given according to \citet{Hacar2013}, H2013, and \citet{Seo2015}, S2015. The given coordinates are the central positions of the maps. The only protostellar core is indicated with an asterisk (*). The region names are given according to \citet{Barnard1927}.}\label{tab:sources}
\centering
\begin{tabular}{lcccr}
\hline\hline
\multicolumn{2}{c}{Core} & $\alpha_{J2000}$ & $\delta_{J2000}$ & Region \\
H2013 & S2015 & ($^h$ $^m$ $^s$) & ($^{\circ}$ $^{\prime}$ $^{\prime\prime}$) & \\
\hline
1 & 12, 13, 14 & 04:17:42.347 & 28:07:30.88 & B10 \\
6 & 8, 9 & 04:18:06.379 & 28:05:34.87 & B10 \\
7 & 19 & 04:18:11.343 & 27:35:33.07 & B211 \\
10 & 22 & 04:19:36.768 & 27:15:32.00 & B213 \\
11* & 23* & 04:19:42.154 & 27:13:31.03 & B213 \\
16 & 33 & 04:21:20.595 & 27:00:13.63 & B213 \\
-- & 35 & 04:24:20.600 & 26:36:02.00 & B216 \\
\hline
\end{tabular}
\end{table}

\begin{table*}
\caption{Methanol lines observed in this study. 
}\label{tab:lines}      
\centering                          
\begin{tabular}{lccccccccr}       
\hline\hline                
Transition & Frequency$^{a}$ & $E_{\rm up}/k^{a}$ & $A^{a}$ & $n_{\rm crit}^{c}$ & $F_{\rm eff}$ & $B_{\rm eff}$ & $\Delta v_{\rm res}$ & rms in $T_{\rm mb}$ & $T_{\rm sys}$\\  
 & (GHz) & (K) & ($10^{-5}$~s$^{-1}$) & (10$^5$~cm$^{-3}$) & & & (km~s$^{-1}$) & (K) & (K) \\   
\hline                        
(2$_{1,2}$--$1_{1,1}$)-$E_2$ & 96.739362 & 12.53$^{b}$ & 0.2558 & 0.82 & 0.95 & 0.80 & 0.06 & 0.08--0.17 & 70--100 \\   
(2$_{0,2}$--$1_{0,1}$)-$A^+$ & 96.741375 & 6.96  & 0.3408 & 1.09 & 0.95 & 0.80 & 0.06 & 0.08--0.17 & 70--100 \\
(3$_{1,3}$--$2_{1,2}$)-$E_2^d$ & 145.097370 & 19.51 & 1.0957 & 2.74 & 0.93 & 0.73 & 0.04 & 0.06--0.14 & 100--160 \\
(3$_{0,3}$--$2_{0,2}$)-$A^+$ & 145.103152 & 13.93 & 1.2323 & 13.2 & 0.93 & 0.73 & 0.04 & 0.06--0.14 & 100--160 \\
\hline
(2$_{0,2}$--$1_{0,1}$)-$E_1$ & 96.744550 & 20.08$^{b}$ & 0.3407 & 1.09 & 0.95 & 0.80 & 0.06 & 0.08--0.17 & 70--100 \\
(3$_{0,3}$--$2_{0,2}$)-$E_1$ & 145.093760 & 27.1 & 1.2314 & 103 & 0.93 & 0.73 & 0.04$^{e}$ & 0.06--0.14 & 100--160\\
\hline                                   
\end{tabular}\\
\begin{flushleft}
Notes.
$^{a}$The methanol 3~mm and 2~mm frequencies, 3~mm energies and Einstein coefficients are taken from \citet{Bizzocchi2014} following \citet{Xu1997} and \citet{Lees1968}, also available at the Jet Propulsion Laboratory (JPL) database \citep{Pickett1998}. $^{b}$Energy relative to the ground 0$_{0,0}$, A rotational state. $^{c}$The critical densities are calculated for kinetic temperature of 10~K with an assumption of optically thin lines. $^{d}$The parameters for the 2~mm methanol lines are taken from the JPL database \citep[frequencies and energies,][]{Pickett1998} and Leiden Atomic and Molecular Database, LAMDA, \citep[Einstein coefficients,][]{Schoeier2005}. $^{e}$The 2~mm methanol lines for cores Hacar01, Hacar11, and Seo35 were observed with the Fast Fourier Transform Spectrometer FTS~50 backend with spectral resolution of 50~kHz or 0.10~km~s$^{-1}$. The transitions with $E_{\rm up}/k>20$~K (below the horizontal line) were not detected in the majority of the cores and were not used in the analysis (see Sect.~\ref{sec:app_co} for details). 
\end{flushleft}
\end{table*}

\begin{figure*}
\includegraphics[height=7.5cm,keepaspectratio]{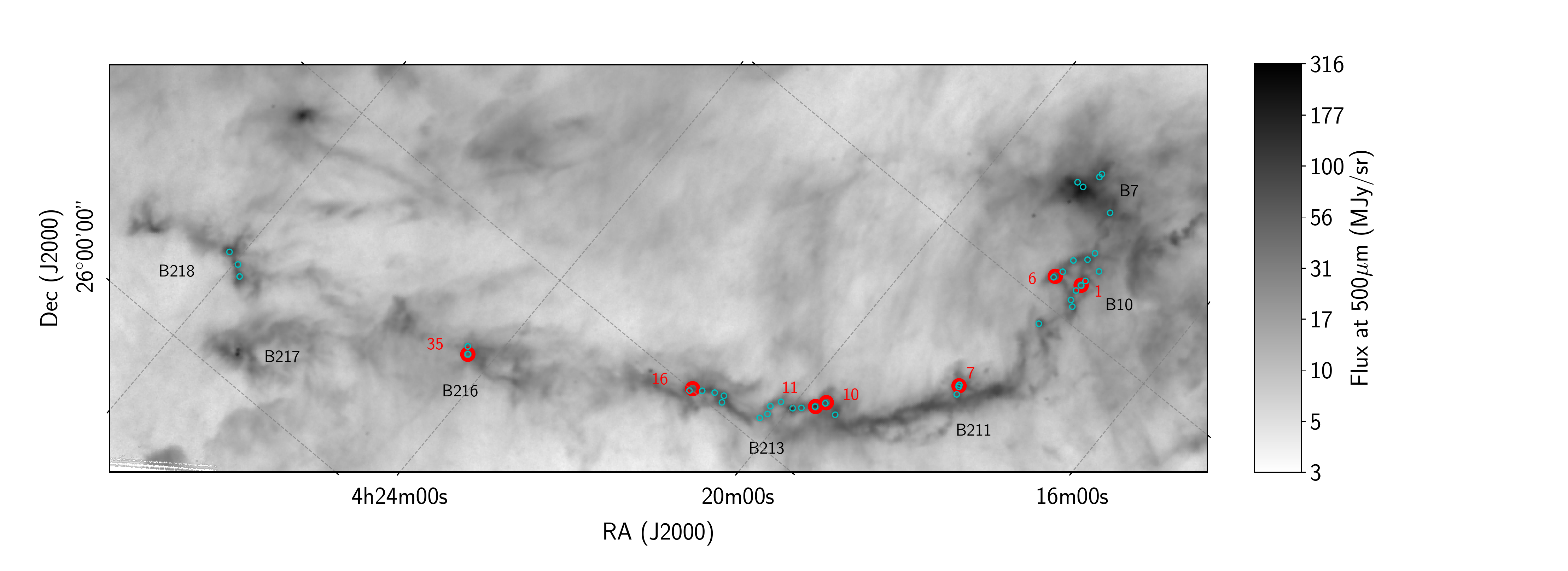}
\caption{500~$\mu$m dust continuum emission toward the L1495 filament of the Taurus molecular cloud mapped by {\it Herschel}/SPIRE \citep{Palmeirim2013}. The cyan circles show the 31 cores observed in methanol by \citet{Scibelli2020}. Red numbered dots show the dense cores mapped in methanol in this work. We refer to the cores in this work according to these numbers (see Table~\ref{tab:sources}).}\label{fig:L1495}
\end{figure*}

We mapped six methanol lines at 96.7~GHz and 145.1~GHz toward seven dense cores of the L1495 filamentary structure (see Fig.~\ref{fig:L1495} and Tables~\ref{tab:sources} and~\ref{tab:lines}) with the IRAM 30~m telescope (IRAM projects 013-18, 125-18, and 031-19). The observations were performed on 2018 October 17--23, 2019 March 27--29 and September 16 under acceptable weather conditions, with precipitable water vapour $pwv$=1--10~mm. The on-the-fly maps were obtained with the Eight MIxer Receivers EMIR~090 (3~mm band) and EMIR~150 (2~mm band) heterodyne receivers\footnote{\url{http://www.iram.es/IRAMES/mainWiki/EmirforAstronomers}} in position switching mode, and the VESPA (VErsatile SPectrometer Assembly) backend. The spectral resolution was 20~kHz, the corresponding velocity resolutions were $\simeq$0.06~km~s$^{-1}$ for the 3~mm band and $\simeq$0.04~km~s$^{-1}$ for the 2~mm. The beam sizes were $\simeq$26$^{\prime\prime}$ for the 3~mm band and $\simeq$17$^{\prime\prime}$ for the 2~mm. The system temperatures were 90--627~K depending on the frequency. 

The exact line frequencies, beam efficiencies, beam sizes, spectral resolutions and sensitivities are given in Table~\ref{tab:lines}. Sky calibrations were obtained every 10--15~minutes. Reference positions were chosen individually for each core to make sure that the positions were free of any methanol emission. Pointing was checked by observing QSO~B0316+413, QSO~B0439+360, QSO~B0605-085, Uranus, Mars or Venus every 2~hours and focus was checked by observing QSO~B0439+360, Uranus, Mars or Venus every 6 hours. 

The data reduction up to the stage of the convolved spectral data cubes was performed with the GILDAS/CLASS package\footnote{Continuum and Line Analysis Single-Dish Software \url{http://www.iram.fr/IRAMFR/GILDAS}}. All data were convolved to the 26.8$^{\prime\prime}$ beam with the 8.7$^{\prime\prime}$ pixel, that is consistent with Nyquist sampling, for consistency of the data set. The following spectral analysis was performed with the Pyspeckit module of Python \citep{Ginsburg2011}.

Four out of six observed lines were detected with $S/N\!>\!5$ toward all seven cores. Two lines with the highest energy of the upper level ($E_{\rm up}\simeq20$~K) were detected only toward the brightest areas of core~1. Therefore these two lines were not used for the analysis and only used to test our method of column density measurement (see Sect.~\ref{sec:col_den} and Appendix~\ref{app:col_den}).

\section{Results of observations} \label{sec:results}

To test the model predictions, we apply the most widely used approaches to estimate $A_V$, CO depletion, and methanol abundance and use the most common data so our results would be easily compared to other works. We chose to use the {\it Herschel}\footnote{{\it Herschel} is an ESA space observatory with science instrumentsprovided  by  European-led  Principal  Investigator  consortia  andwith important participation from NASA.} dust continuum emission to measure the molecular hydrogen column density $N({\rm H_2})$ along with the conversion factor connecting $N({\rm H_2})$ and $A_V$, since the {\it Herschel} survey covered the majority of the selected molecular clouds. To measure the methanol column densities we chose the brightest 2~and 3~mm methanol lines. Based on the total nuclear spin of the hydrogen atoms in the methyl group, methanol can take $A$- and $E$-form. Once formed, $A$- and $E$-methanol molecules keep the hydrogen spines. The ratio of $A\!:\!E$ methanol depends on the temperature at which methanol was formed, with $A\!:\!E\!=\!1\!:\!1$ at 30--40~K, increasing at lower temperatures \citep[see e.g.,][]{Wirstrom2011}. At the temperatures of our cores, 10--12~K, the $A\!:\!E$ ratio should be $\simeq\!1.3$. In our observational set, we have only two lines of each methanol form, which is not enough to measure column densities via rotational diagrams robustly, so we combine them and assume the 1:1 $A\!:\!E$ methanol ratio. Besides that, \citet{Bizzocchi2014} found $A\!:\!E\!=\!1.03\!\pm\!0.26$ in another Taurus core, L1544. To measure the column densities of both methanol and CO we assumed the lines are consistent with local thermodynamic equilibrium (LTE) and optically thin emission (which is justified, see Sect.~\ref{sec:co} and~\ref{sec:col_den} for details). We also use a large homogeneous data set of typical dense cores in a low-mass star-forming region. To measure $T_{\rm dust}$ and $N({\rm H_2})$ we used the data from {\it Herschel} Spectral and Photometric Imaging Receiver, SPIRE, \citep{Palmeirim2013}. To trace CO depletion, we used the C$^{18}$O(2--1) observations by \citet{Tafalla2015}. We then used $N({\rm H_2})$ to measure the CO and methanol abundances. These steps are described in the following section.
\subsection{H$_2$ column density and $A_V$}

The {\it Herschel} space telescope carried out the most extensive coverage of dust continuum emission of the molecular clouds in our Galaxy, with PACS\footnote{Photodetector Array Camera \& Spectrometer.} and SPIRE instruments. The {\it Herschel} science archive is the main source now for the dust continuum emission data in submillimeter range to estimate dust temperature and molecular hydrogen column densities in molecular clouds \citep[see e.g.,][]{Andre2014,Friesen2017,Ladjelate2020}. 

We use the archive {\it Herschel}/SPIRE 250, 350, and 500~$\mu$m dust continuum emission map \citep{Palmeirim2013} downloaded from the {\it Herschel} Science Archive (Observation ID 1342202254) to measure molecular hydrogen column density $N$(H$_2$), visual extinction $A_V$, and dust temperature $T_{\rm dust}$. We smooth the 250 and 350~$\mu$m maps to the largest beam (of 500~$\mu$m) of 38$^{\prime\prime}$ and fit a modified black body function. We use emissivity index $\beta=2.0$, typical for starless cores \citep{Draine2007,Schnee2010}, close to the value $\beta=2.4$ previously found toward pre-stellar cores L1544 and Miz-2 in Taurus \citep{Chacon-Tanarro2017,Bracco2017}. We use dust opacity $\kappa_{250 \mu m}$=0.144~cm$^2$~g$^{-1}$ based on $\kappa_{\nu}$=0.1($\nu$/10$^{12}$[Hz])$^{\beta}$~cm$^2$~g$^{-1}$ \citep{Beckwith1990}.

To measure visual extinction $A_V$, we scale H$_2$ column density following the conversion factor given in \citet{Guver2009}:
\begin{equation}\label{rel1}
A_V = \frac{N{\rm (H_2)[ cm^{-2}]}}{1.105\times10^{21}[{\rm cm^{-2}/mag}]}.
\end{equation}

Our peak H$_2$ column density values agree within the errors or differ by 20\% with those measured by \citet{Seo2015} using the 500~$\mu$m {\it Herschel}/SPIRE data only (see the comparison of the peak values in Table~\ref{tab:nh2}). To evaluate our H$_2$ column densities with an independent method, we scaled the $A_V$ map of L1495 by \citet{Schmalzl2010} to $N({\rm H_2})$ using the relation~(\ref{rel1}). In general, the column densities agree within 20\% and differ by a factor of two in cores 11 and 35. The difference is not systematic (see the peak values in Table~\ref{tab:nh2} and comparison of all available points in Fig.~\ref{fig:herschel-vs-av}). The $A_V$ map by \citet{Schmalzl2010} was obtained by infrared photometric observations of background stars, so it suffers from gaps and interpolation effects, for example, there are two gaps covering a large area of core~35 and a gap toward core~11, those might have led to the factor of 2 difference in the estimated $N$(H$_2$). Thus we chose to use the {\it Herschel} data to measure both $N({\rm H_2})$ and $A_V$.

\begin{table}
    \caption{Peak column and volume densities of H$_2$ toward the studied cores.}
    \label{tab:nh2}
    \centering
    \begin{tabular}{ccccccccc}\hline\hline
     Core & \multicolumn{3}{c}{$N({\rm H_2})$ (10$^{22}$~cm$^{-2}$)} & \multicolumn{5}{c}{$n({\rm H_2})$ (10$^{5}$~cm$^{-3}$)} \\
     \hline
        & SED & S10$^N$ & S15 & & * & S10$^n$ & M14 & WT16 \\
        \hline
     1  & 2.65 & 2.16 & 2.53$^{+0.19}_{-0.07}$ & & 1.50 & -- & 26 & 0.69 \\
     6  & 2.57 & 3.21 & 2.64$^{+0.17}_{-0.27}$ & & 1.39 & -- &  6.8 & 0.81 \\
     7  & 2.40 & -- & 2.21$^{+0.28}_{-0.28}$ & & 0.78 & -- & 1.9 & 0.40 \\
     10 & 2.07 & 2.26 & 1.71$^{+0.15}_{-0.10}$ & & 0.85 & 0.18 & 1.6 & -- \\
     11 & 2.50 & 1.23 & 2.05$^{+0.08}_{-0.07}$ & & 1.83 & 0.18 & -- & -- \\
     16 & 2.80 & 2.80 & 2.38$^{+0.21}_{-0.10}$ & & 1.57 & 0.14 & -- & -- \\
     35 & 1.60 & 3.16 & 1.73$^{+1.59}_{-0.36}$ & & 0.51 & 0.17 & -- & -- \\
     \hline
    \end{tabular}
\begin{flushleft}
Notes. Columns: SED -- our $N({\rm H_2})$ based on modified black-body spectral energy distribution (SED); S10$^N$ -- $N({\rm H_2})$ based on conversion of $A_V$ from \citet{Schmalzl2010}; S15 -- $N({\rm H_2})$ from \citet{Seo2015}; * -- $n({\rm H_2})$ modeled in this work; S10$^n$ -- $n({\rm H_2})$ from \citet{Schmalzl2010}; M14 -- $n({\rm H_2})$ from \citet{Marsh2014}; WT16 -- $n({\rm H_2})$ from \citet{Ward-Thompson2016}.
\end{flushleft}
\end{table}

\subsection{CO depletion factor}\label{sec:co}

In cold, dense, quiescent gas, CO freezes out onto dust grains and partly it is transformed into methanol. The level of CO freeze-out thus is one of the major factors affecting methanol formation \citep[e.g.,][]{Whittet_2011}. It is commonly expressed as a CO depletion factor, $f_d$, that is defined as the ratio of the reference maximum abundance of CO ($X_{\rm ref}({\rm CO})$) in the cloud to the abundance of CO measured in the gas phase in the region of interest,~$X$(CO). CO depletion is also often used as a chemical age indicator that helps to link the observed molecular abundances to the modeled ones \citep[e.g.,][]{Jimenez-Serra2016,Lattanzi2020}. 

To measure the gas-phase CO abundance toward the cores, we use the C$^{18}$O(2--1) maps obtained with the IRAM~30~m antenna, convolved to 23$^{\prime\prime}$ beam, by \citet{Tafalla2015}\footnote{The data are available via Strasbourg astronomical Data Center (CDS): \url{https://cdsarc.cds.unistra.fr/ftp/J/A+A/574/A104/}}, except for core~35 which belongs to the B216 region, there is no published C$^{18}$O(2--1) map for B216. $X{\rm _{\rm ref}(CO)}$ is defined in Sect.~\ref{sec:app_co}.
\subsubsection{C$^{18}$O column densities}

\citet{Tafalla2015} find more than one C$^{18}$O line velocity components toward many positions in their map. Some of these components belong to the dense cores, the others belong to their envelopes or other material in the molecular cloud on the line of sight. \citet{Hacar2013} and \citet{Tafalla2015} resolve and analyze the material as small fibers composing the bigger filament. Since the molecular hydrogen column density measured both via $A_V$ and dust continuum emission do not contain kinematic information, we integrate all emission in the C$^{18}$O(2--1) spectra to measure the integrated intensity $W$ and column density $N_{\rm tot}$. 

To measure the C$^{18}$O column densities, we assume that the C$^{18}$O(2--1) lines are optically thin and consistent with LTE. Following \citet{Caselli2002-b}, the column density then is:
\begin{eqnarray}
    N_{\rm tot}= && \frac{8\pi W}{\lambda^3A_{ul}}\frac{g_l}{g_u}\frac{1}{J_{\nu}(T_{\rm ex})-J_{\nu}(T_{\rm bg})}\frac{1}{1-\exp{(-h\nu/kT_{\rm ex})}} \nonumber \\
    && \times \frac{Q_{\rm rot}}{g_l\exp{(-E_l/kT_{\rm ex})}},
\end{eqnarray}
where $A_{ul}$ is the Einstein coefficient, $g_l$ and $g_u$ are the statistical weights of the upper and lower levels, $\lambda$ is the wavelength, $J_{\nu}(T)$ is the equivalent Rayleigh-Jeans temperature, $T_{\rm bg}$=2.7~K is the cosmic background temperature, $T_{\rm ex}$ is the excitation temperature, $E_l$ is the energy of the lower level. The partition function $Q_{\rm rot}$ of linear molecules (such as CO) is given by
\begin{equation}
    Q_{\rm rot}=\sum_{J=0}^{\infty}(2J+1)\exp{(-E_J/kT)},
\end{equation}
where $J$ is the rotational quantum number, $E_J=J(J+1)hB$, and $B$ is the rotational constant. We assume $T_{\rm ex}$=10~K which is consistent with the gas temperature measured by \citet{Seo2015}. With $T_{\rm ex}$=10~K, equation~2 for the C$^{18}$O(2--1) line can be simplified to $N_{\rm tot}\simeq$6.528$\times$10$^{14} W$, where $W$ is in K~km~s$^{-1}$ and $N_{\rm tot}$ is in cm$^{-2}$.

 The excitation conditions of C$^{18}$O are close to those of the main isotopologue CO. The optically thin critical density of the C$^{18}$O(2--1) transition is $2\times10^4$~cm$^{-3}$, such that at lower densities it would deviate from LTE. With the statistical equilibrium radiative transfer code RADEX, we estimated that only toward the less dense outskirts of the cores ($n\sim10^3$~cm$^{-3}$) $T_{\rm ex}\simeq5$~K, which gives two times higher $N_{\rm tot}$ than that with LTE at 10~K.
 
\subsubsection{CO depletion: choice of the reference CO abundance}\label{sec:app_co}

\begin{figure}
    \centering
    \includegraphics[height=4.5cm,keepaspectratio]{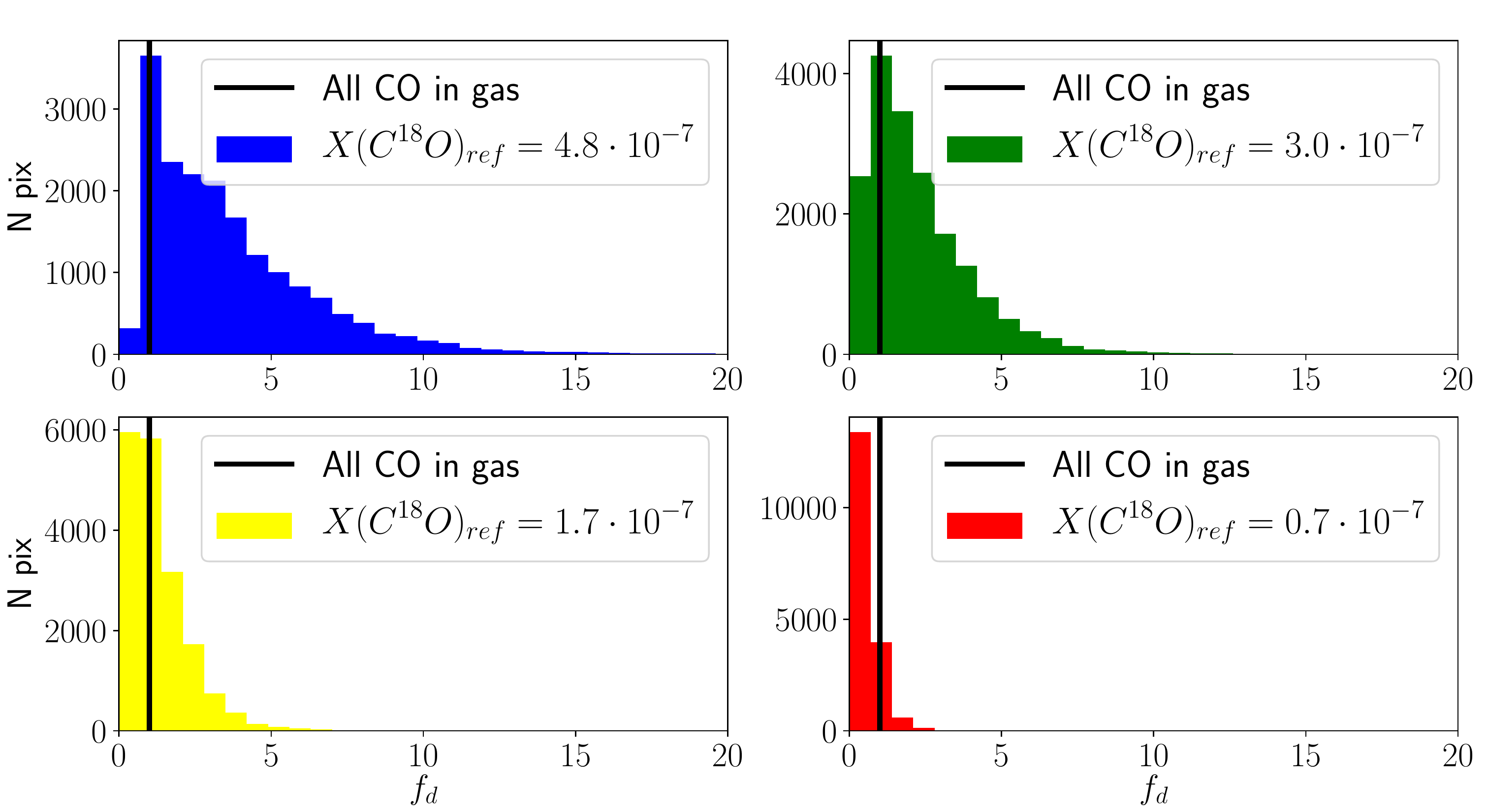}
    \caption{CO depletion $f_d$ measured across all B7, B10, B211, B213 and B218 regions, mapped by \citet{Tafalla2015}, using various reference CO abundance. The black vertical line shows $f_d$=1, when all CO is in the gas phase.}
    \label{fig:fd_cloud}
\end{figure}

The reference abundance of CO with respect to H$_2$ in the local interstellar medium (ISM) has been estimated in a number of studies with a spread between 0.4 and 2.7$\times 10^{-4}$ \citep{Wannier1980,Frerking1982,Lacy1994,Tafalla2004b}. The numbers given in literature are presented in Table~\ref{tab:ref_co}. 

\begin{table*}
    \caption{CO depletion with different $X{\rm (C^{18}O)}$.}
    \label{tab:ref_co}
    \centering
    \begin{tabular}{lccccccll}\hline\hline
    $X$(CO) & $X{\rm (C^{18}O)}$ & \multicolumn{3}{c}{$f_d$}& \multicolumn{2}{c}{$N(f_d<1)$}& citation & also used in \\ 
     & & min & median & max & & & &\\
    \hline
    0.39$\times$10$^{-4}$* & $0.7\times 10^{-7}$ & 0.04 & 0.36 & 5.9 & 15954 & 70\% & \citet{Tafalla2004b} \\
    0.85$\times$10$^{-4}$& $1.7\times 10^{-7}$ & 0.09 & 0.87 & 14.4 & 8639 & 38\% & \citet{Frerking1982} & \citet{Bacmann2002} \\
      &  &   &   &   &   &   & \citet{Alonso-Albi2010} & \citet{Jorgensen2002}\\
      &  &   &   &   &   &   &  & \citet{Crapsi2005}\\
    1.7$\times$10$^{-4}$& $3.0\times 10^{-7 a}$ & 0.17 & 1.54 & 25.5 & 4675 & 20\% & \citet{Lacy2017}\\
    2.7$^{+6.4}_{-1.2}\times10^{-4}$& $4.8\times 10^{-7 a}$ & 0.27 & 2.47 & 40.8 & 1625 & 7\%  & \citet{Lacy1994} & \citet{Lee2003} \\
    \hline
    \end{tabular}
    \begin{flushleft}
    $^a$Abundances found applying the fractional abundance $^{16}$O/$^{18}$O=560 from \citet{Wilson1994}.
    \end{flushleft}
\end{table*}

To choose the reference abundance of CO from the values given in Table~\ref{tab:ref_co}, we calculated CO depletion factors for each data point of the entire C$^{18}$O(2--1) map of L1495 filament provided in \citet{Tafalla2015} (B7, B10, B211, B213, B218). The area mapped by \citet{Tafalla2015} reaches the relatively diffuse outskirts of the filament, where CO should be undepleted. By definition, the depletion factor must be equal to unity there: $f_d=1$. Thus, our goal is to find the reference abundance of CO that provides the value of $f_{d}$ close to unity in the diffuse outskirts of the filament, and also ensures the minimum number of data points in the C$^{18}$O(2--1) map of L1495 with nonphysical values of $f_d<1$. To estimate $f_d$, we use only well-defined column densities, with $N({\rm C^{18}O})/\Delta N({\rm C^{18}O})>5$. The resulting distributions of depletion factors across the entire L1495 map obtained for different reference abundances of CO are presented in Fig.~\ref{fig:fd_cloud}. The chosen reference fractional abundances affect the CO depletion factor, with the highest $X_{\rm ref}$(CO) giving the largest $f_d$ and the widest range of $f_d$ (see Table~\ref{tab:ref_co}). Taking into account the uncertainties of the calculated C$^{18}$O column densities (including the LTE assumption, that results in underestimation of the column density), the first two reference abundances give median $f_d$ values in the cloud below 1, which is nonphysical. Toward the least depleted areas, we find $f_d\simeq0.2$, 0.5, 0.9, and 1.5 with $X_{\rm ref}({\rm C^{18}O})=0.7\times 10^{-7}$, $1.7\times10^{-7}$, $3.0\times10^{-7}$, and $4.82\times10^{-7}$, respectively. The choice of the reference value was based on the one which gave the lowest number of points (out of the total 22941) with $f_d<1$.

Thus we use the highest reference abundance of CO equal to 2.7$\times$10$^{-4}$ w.r.t. H$_{2}$ (which gives 4.8$\times$10$^{-7}$ of C$^{18}$O w.r.t. H$_{2}$) that gives only 7\% of nonphysical $f_d$ values across the maps, to measure CO depletion. These 7\% may be caused by a large spread of this reference value \citep[see Table~\ref{tab:ref_co} and][]{Lacy1994}. The number of the nonphysical values is small, and we do not aim at fitting the best reference CO abundance, we just use one of the well established reference values that fits better.

\subsubsection{C$^{18}$O abundance and CO-depletion factor}

\begin{figure}
\includegraphics[height=5.5cm,keepaspectratio]{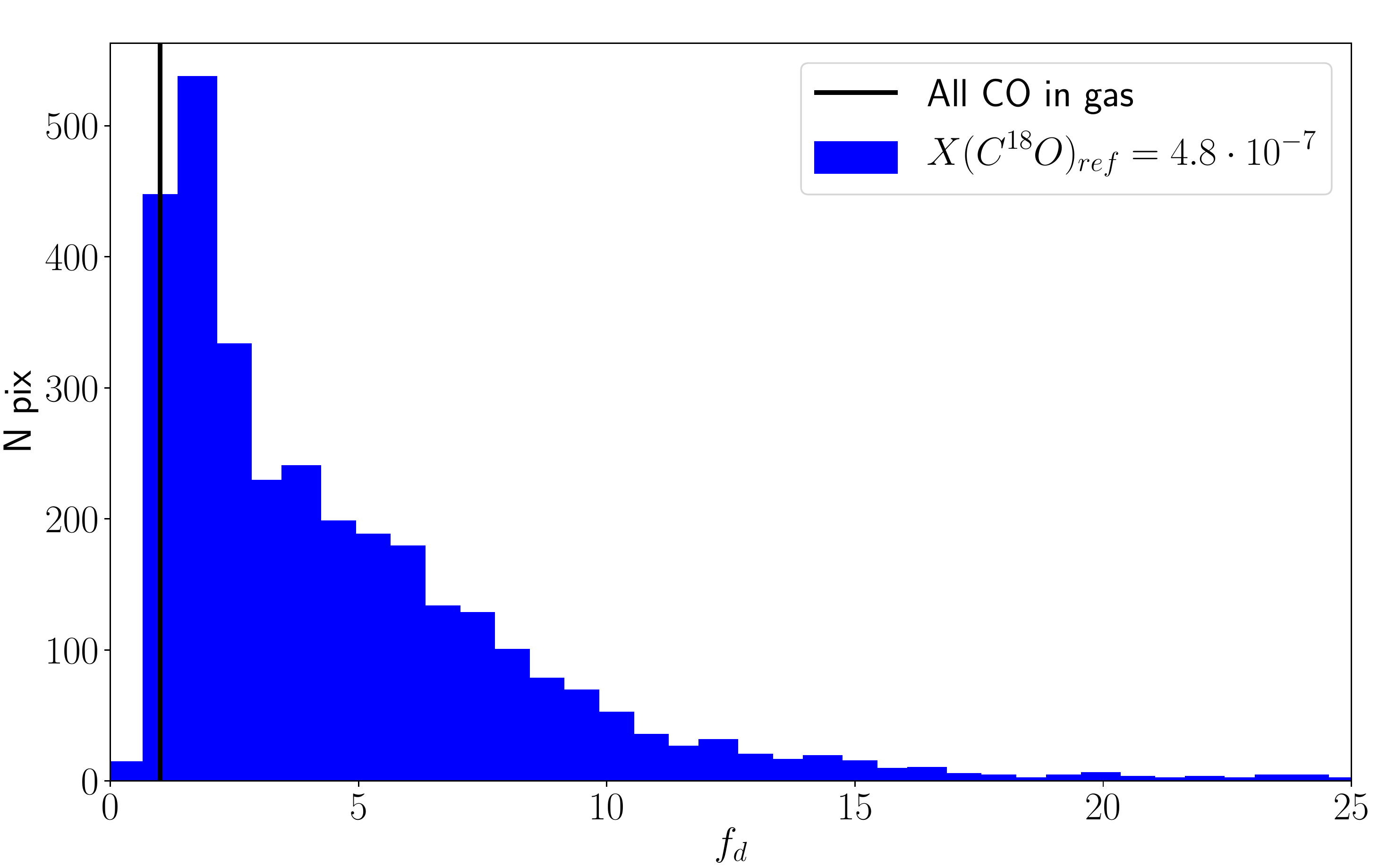}
\caption{The distribution of CO depletion factor within the maps of all studied cores (except for core~35). The black vertical line represents $f_d$=1 (all CO in gas phase).
}\label{fig:fd-hist}
\end{figure}

\begin{figure*}
    \centering
    \includegraphics[height=9.0cm,keepaspectratio]{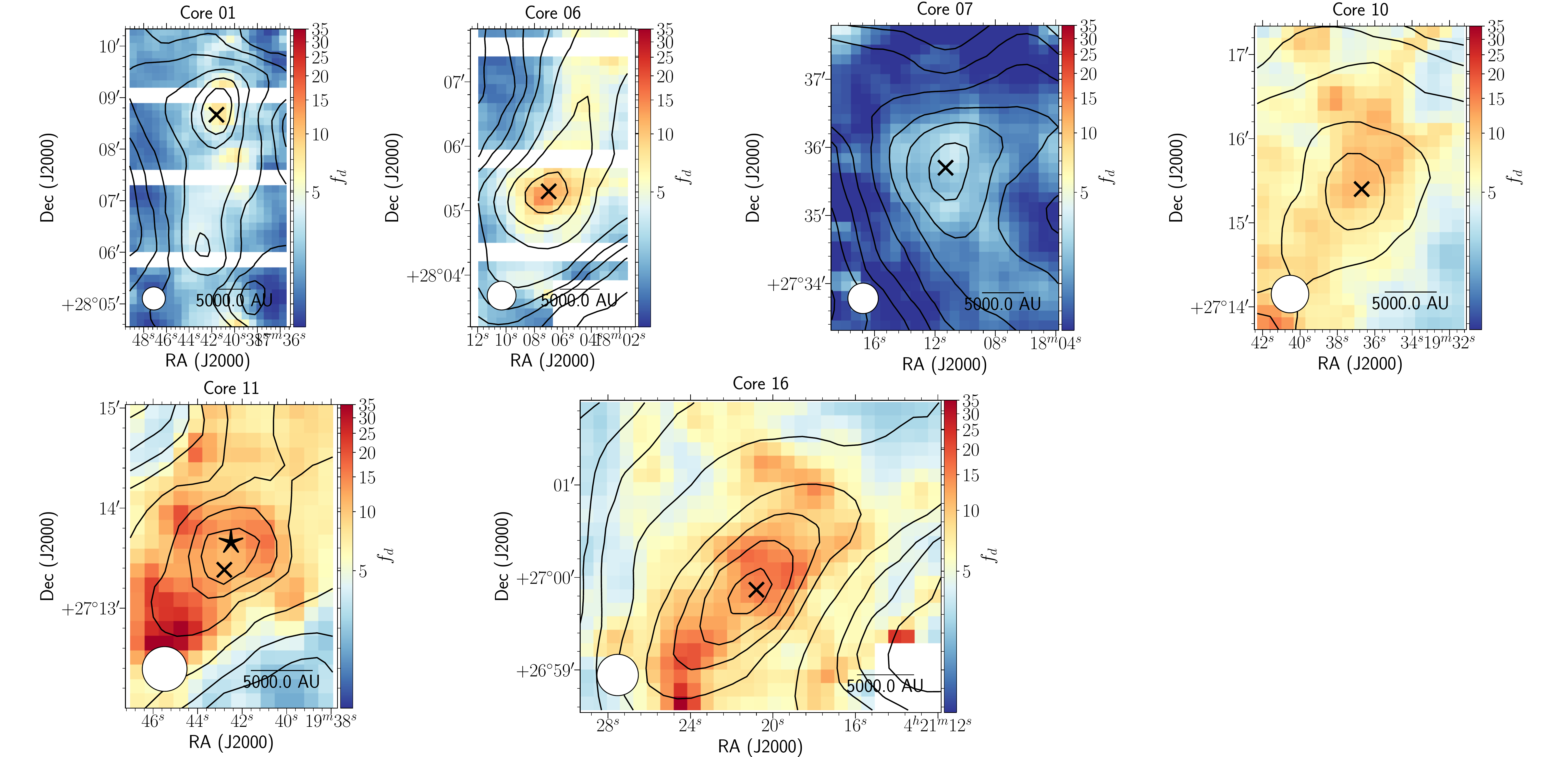}
    \caption{CO depletion factor (color scale) maps toward the observed cores based on C$^{18}$O observations from \citet{Tafalla2015}. Black contours show visual extinction at $A_V=$3, 4, 5, 6, 8, 12, 16, 20, 24~mag. The top $A_V$ contours are at $A_V$=24~mag for core~16 and at $A_V$=20~mag for the other cores. The black star shows the position of Class 0 protostar IRAS~04166+2706 \citep{Santiago-Garcia2009}, crosses show the {\it Herschel}~/~SPIRE dust emission peaks. White circle on the bottom left of each map shows the 23$^{\prime\prime}$ beam. The white strips in the maps of cores 1 and 6 originate from the NaNs in the B10 map available at CDS.}
    \label{fig:fd_map}
\end{figure*}

The depletion factor of CO is generally moderate in L1495, ranging from 1 to 44 toward the studied cores (see Fig.~\ref{fig:fd-hist} and~\ref{fig:fd_map}; there are just a few pixels with high $f_d$, so we limit the histogram in Fig.~\ref{fig:fd-hist} at $f_d$=25 to show the dynamic range better). Among them, core~7 shows the lowest depletion factor, $f_d$=1--4, as well as the entire B211 region with $f_d$=1--4. Cores 10, 11, and 16 (B213 region) show very similar depletion values, 2--44, the highest in our data set (that confirms B213 being more dynamically evolved than B211 and B10), with the median of $f_d$=6--8, high values both for starless and protostellar core. Cores 1 and 6 of the B10 region show intermediate values of $f_d$=1--14. Given the angular resolution of {\it Herschel} maps, these numbers could be lower limits of $f_d$.  CO depletion toward core~35 was not measured since there are no available CO maps for this region.

\subsection{Methanol distribution}\label{sec:distrib}

\begin{figure*}
    \centering
    \includegraphics[height=9.0cm,keepaspectratio]{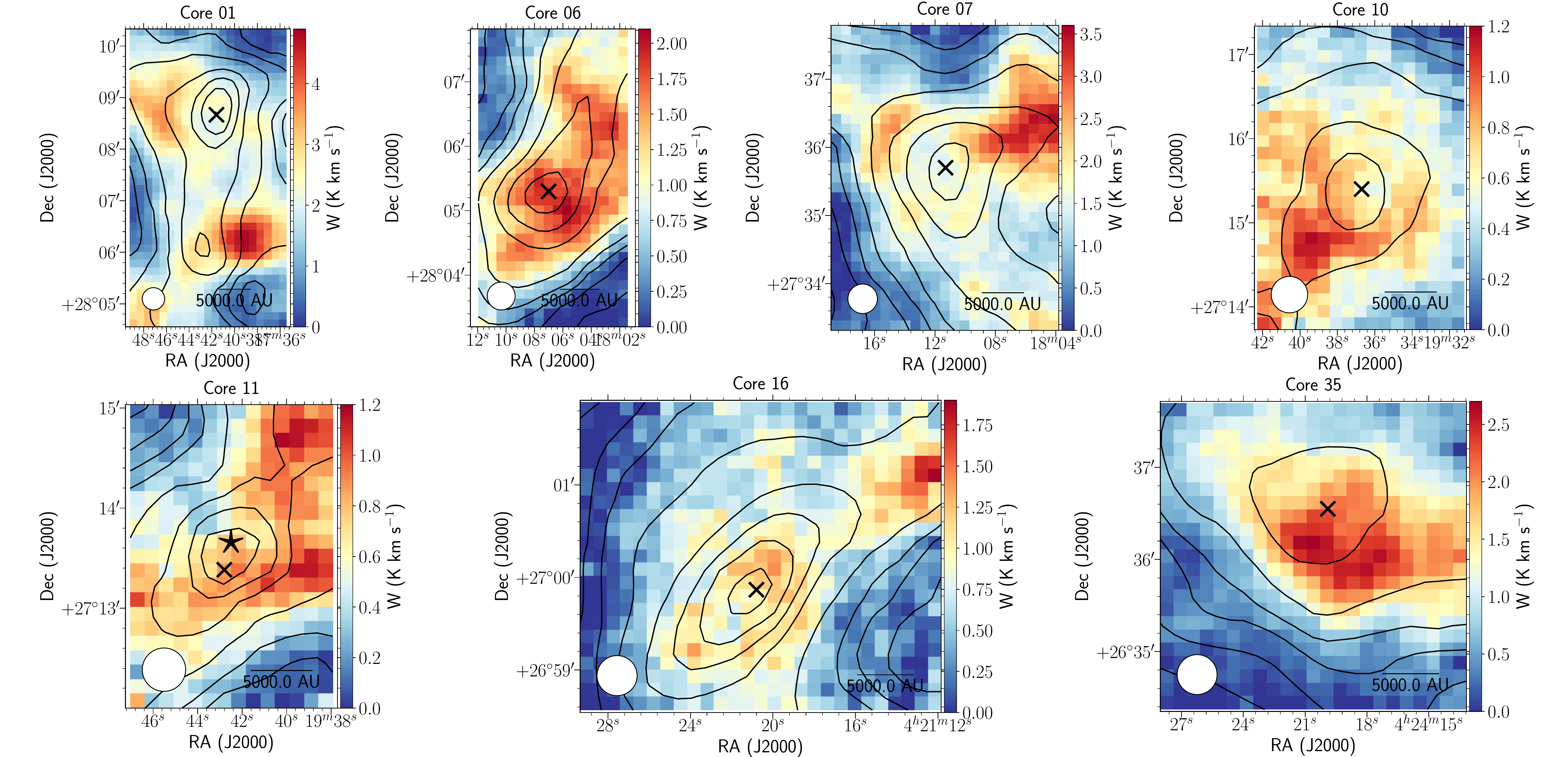}
    \caption{Integrated intensity of all observed methanol lines toward the observed cores (color scale) and visual extinction (black contours at $A_V=$3, 4, 5, 6, 8, 12, 16, 20, 24~mag). The top $A_V$ contours are at $A_V$=24~mag for core~16 and at $A_V$=20~mag for the other cores. Black star shows the position of Class 0 protostar IRAS~04166+2706 \citep{Santiago-Garcia2009}, crosses show the {\it Herschel}~/~SPIRE dust emission peaks. White circle on the bottom left of each map shows the 26$^{\prime\prime}$ IRAM beam.}\label{fig:maps}
\end{figure*}

Figure~\ref{fig:maps} shows the distribution of integrated intensity over all observed methanol lines and contours of the dust continuum emission mapped with {\it Herschel}/SPIRE \citep{Palmeirim2013} toward all seven observed cores. The gas phase methanol is expected to be most abundant in the areas where CO is actively freezing onto the dust grains, and the methanol production overcomes its freeze-out \citep[see][]{Vasyunin2017}. We assume the lines are optically thin (see Sect.~\ref{sec:col_den}), so the methanol distribution can be traced by the intensity maps.

Toward three cores methanol appears in a shell around the densest area (cores 1, 7, and~10) where it is moderately (core~10) or significantly (cores~1 and~7) depleted. Toward two cores (6 and~16), methanol emission peaks are close to the dust emission peaks. Cores 11 and 35 show something in between: the methanol peaks are close to the dust emission peak, however the highest contour of methanol emission has an elongated arch-like shape and rather surrounds the highest dust emission contour. There are no CO observations for core~35. 

The methanol emission in the methanol-rich shells around the cores 1, 7, 10, 11, and 35 is not uniform, with one or two spots of enhanced methanol toward each core. Such methanol distribution (in a shell around the dense parts with one or two emission peaks within the shell) was observed before toward other dense cores in Taurus \citep[L1498, L1544, L1521E;][]{Tafalla2006,Bizzocchi2014,Nagy2019} and Ophiuchus \citep[Oph-H-MM1;][]{Harju2020}. We confirm that shell-like methanol distribution is also widespread in the dense cores of the L1495 filament.  

\subsection{Methanol column densities}\label{sec:col_den}

We measure the methanol column densities via the rotational diagrams based on the four brightest lines detected across all cores. We use only the pixels with all four lines detected with signal-to-noise ratio $S/N\!>\!5$. We assume that the methanol lines are optically thin \citep[as is shown in][]{Scibelli2020}, consistent with LTE, and the fractional abundance of $A-$ and $E-$methanol is 1:1. We calculate the column density of the upper level population, $N_{\rm up}$, as
\begin{equation}
N_{\rm up}=\frac{8\pi k W \nu^2}{A h c^3},
\end{equation}
where $k$ is the Boltzmann constant, $W$ is the integrated intensity of the line, $\nu$ is the frequency, $A$ is the Einstein coefficient (given in Table~\ref{tab:lines}), $h$ is the Planck constant, and $c$ is the speed of light \citep[e.g.,][]{Goldsmith1999}.

We fit a linear function $y=ax+b$ to the plot $\ln(N_{\rm up}/g_{\rm up})$ vs $E_{\rm up}/k$, where $g_{\rm up}$ is the statistical weight of the upper level ($g_J=2J+1$, with $J$ being the rotational quantum number), $E_{\rm up}$ is the energy of the upper level. Then the rotational temperature is $T_{\rm rot}=-1/a$, and the total column density is $N_{\rm tot}=e^bQ_{\rm rot}$, where $Q_{\rm rot}$ is the rotational partition function. For an asymmetric rotor (such as methanol) the partition function can be approximated as
\begin{equation}
    Q_{\rm rot}=\frac{5.34\times10^6}{\sigma}\sqrt{\frac{T_{\rm rot}^3}{ABC}},
\end{equation}
where $\sigma=1$ is a symmetry number, $A$, $B$, and $C$ are the rotational constants \citep[in MHz, taken from JPL;][]{Pickett1998}; for details see \citet{Gordy1970}. The uncertainty of the column density was calculated via propagation of errors of the $b$ coefficient. After measuring $T_{\rm rot}$ and $N_{\rm tot}$ we exclude all data points with relative uncertainties higher than 1/3 from our analysis.

The rotational temperature of methanol varies in the range of 6--16~K, with a typical uncertainty of 5--10\% (see Fig.~\ref{fig:t_rot} for the $T_{\rm rot}$ and $\Delta T_{\rm rot}$ distributions and Fig.~\ref{fig:t_rot_maps} for the $T_{\rm rot}$ maps). The median $T_{\rm rot}$=9.1~K is consistent with the gas temperatures of the cores (8--10~K) measured with ammonia by \citet{Seo2015}, that also supports our assumption of LTE.

The total methanol column density varies in the range of (0.5--7.0)$\times$10$^{13}$~cm$^{-2}$ across all cores with typical uncertainty of 10--20\% (see Fig.~\ref{fig:n_tot}). Our column densities measured with the rotational diagrams are consistent with those measured with RADEX by \citet{Scibelli2020} toward the ammonia peaks. Figure~\ref{fig:n_tot_maps} shows the column density maps. Since the brightest methanol emission avoids the dense core centers, we register higher column density values (up to 7$\times$10$^{13}$~cm$^{-2}$) than \citet{Scibelli2020} did (up to 3.5$\times$10$^{13}$~cm$^{-2}$), as they pointed only toward the ammonia peaks. 

Since we assume the lines are optically thin, the column density distribution is similar to the distribution of the integrated intensity (see Fig.~\ref{fig:n_tot_maps}, although toward cores 11 and 16 we have only few points detected with $S/N\!>\!5$ in all four lines).

\subsection{Methanol abundance}

\begin{figure*}
    \centering
    \includegraphics[height=8.5cm,keepaspectratio]{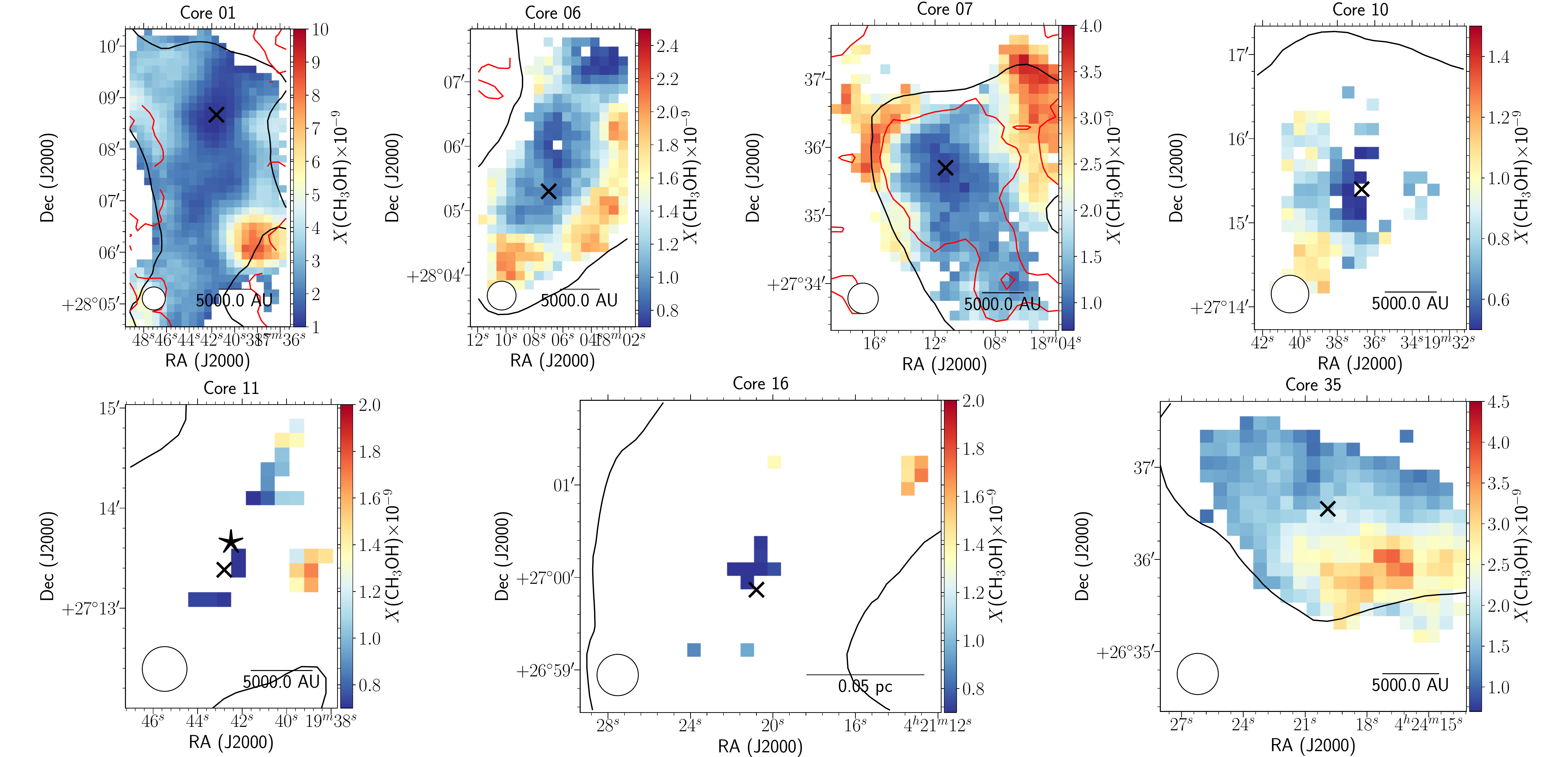}
    \caption{Methanol abundance maps. The red contours show $f_d$=1.6, the black contours show $A_V$=6~mag, the values of observed CO depletion and visual extinction, associated with the highest methanol abundance.}
    \label{fig:x_meth_maps}
\end{figure*}

\begin{figure*}
    \centering
    \includegraphics[height=6cm,keepaspectratio]{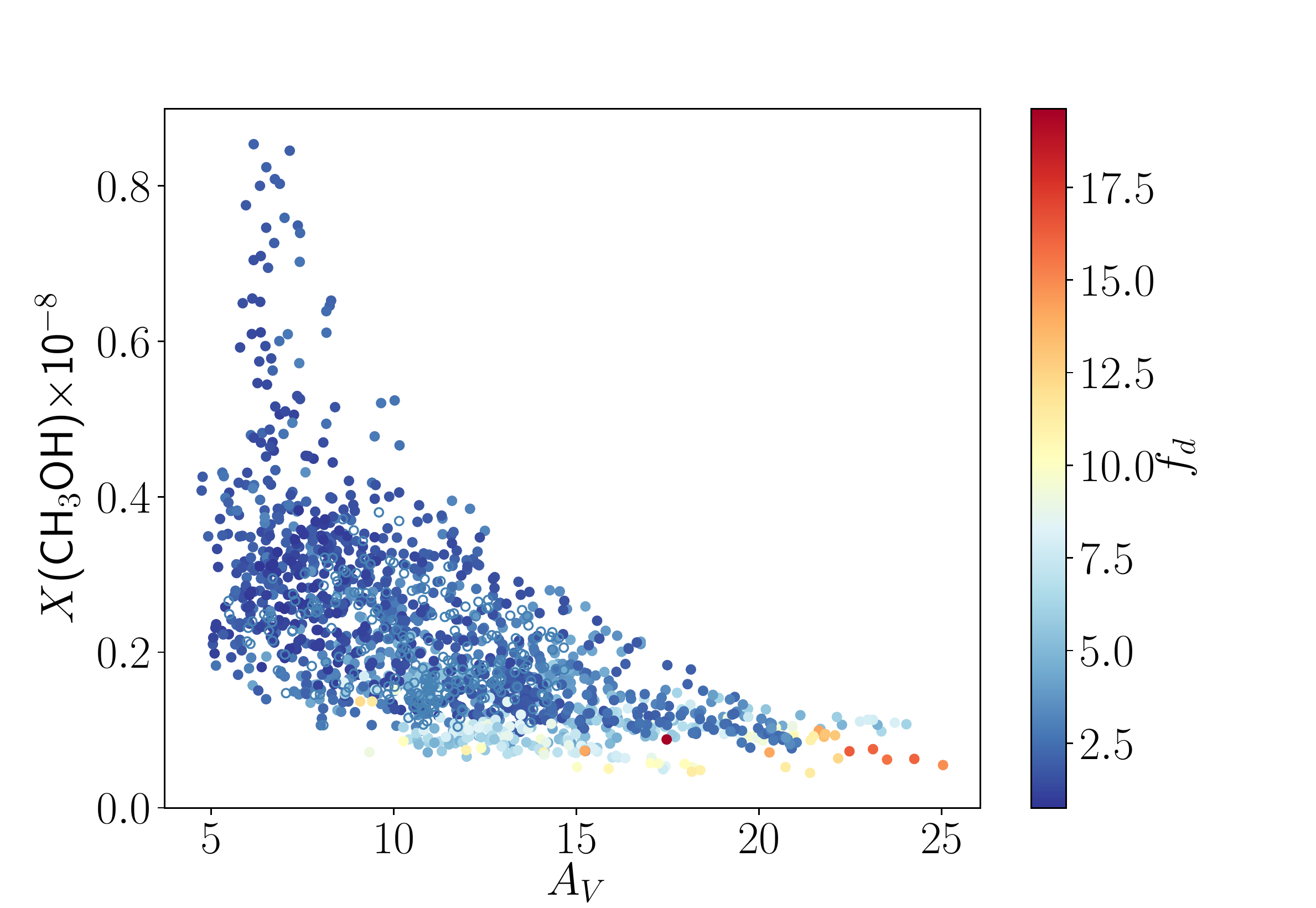}
    \includegraphics[height=6cm,keepaspectratio]{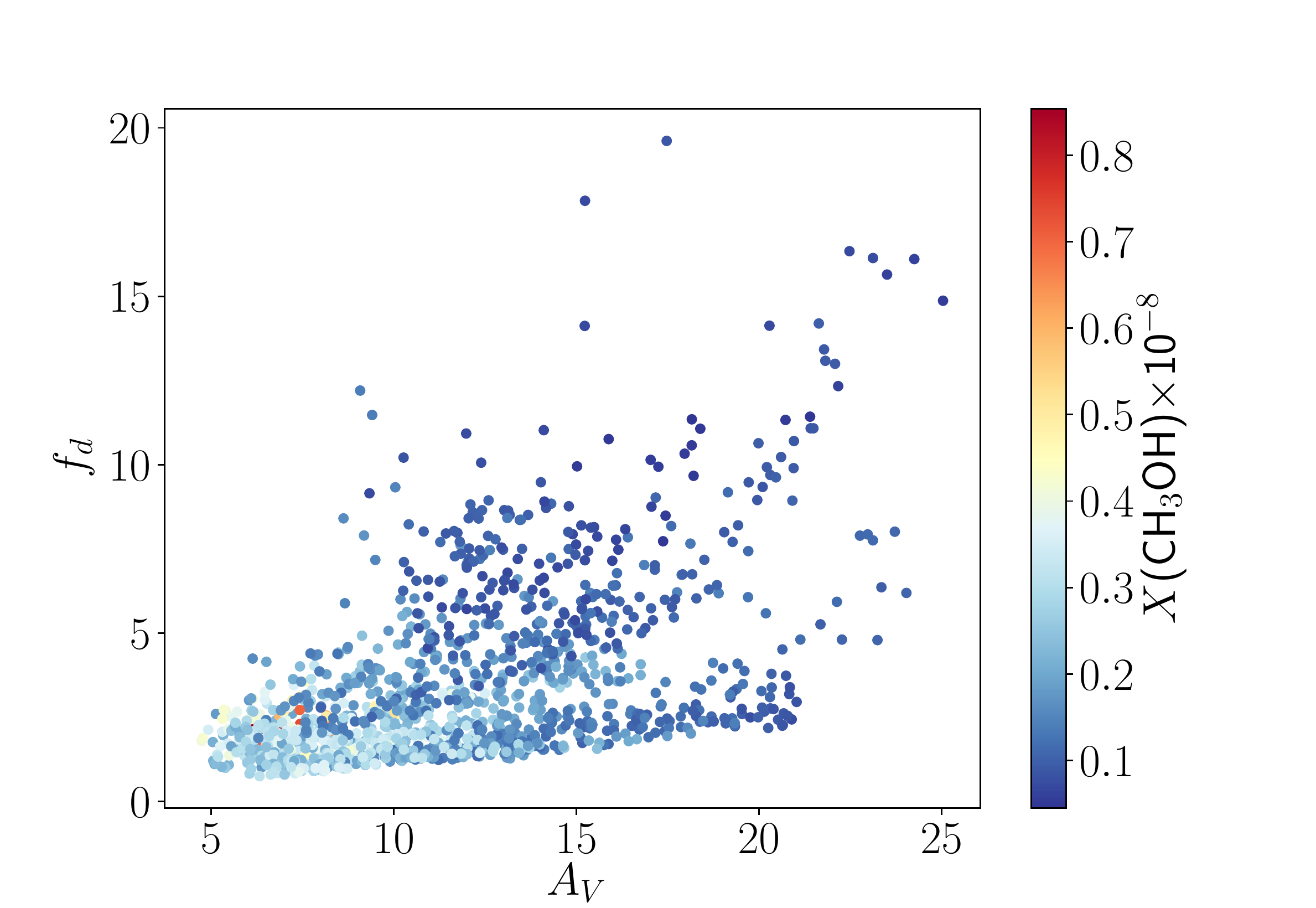}
    \caption{Left: the observed abundance of CH$_3$OH for all of the pixels in all of the maps combined as a function of line-of-sight visual extinction; the color scale represents the CO depletion factor. Right: the CO depletion factor as a function of line-of-sight visual extinction; the color scale represents the methanol abundance.}
    \label{fig:meth_vs_Av}
\end{figure*}

\begin{figure*}\centering
\includegraphics[height=6cm,keepaspectratio]{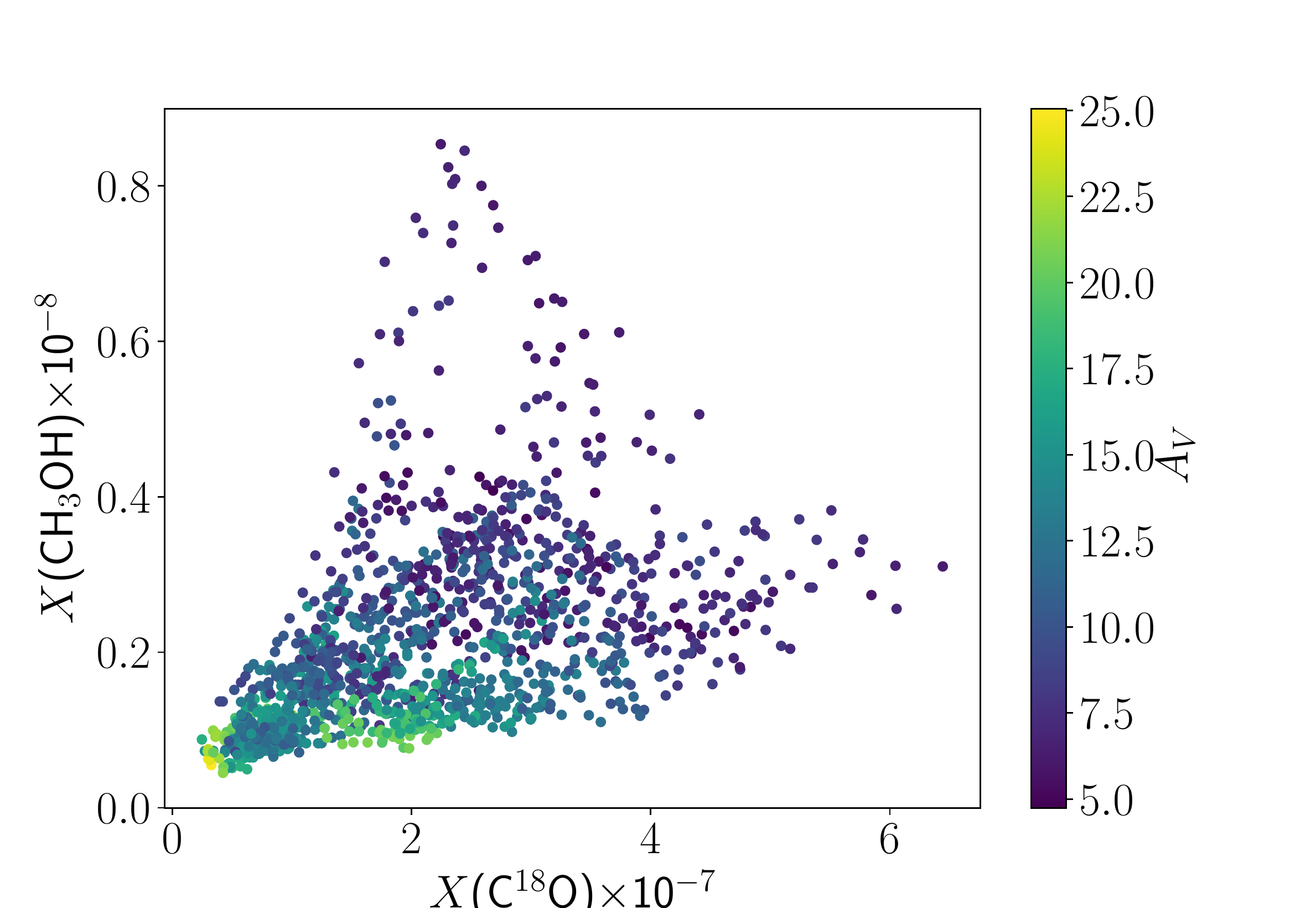}
\includegraphics[height=6cm,keepaspectratio]{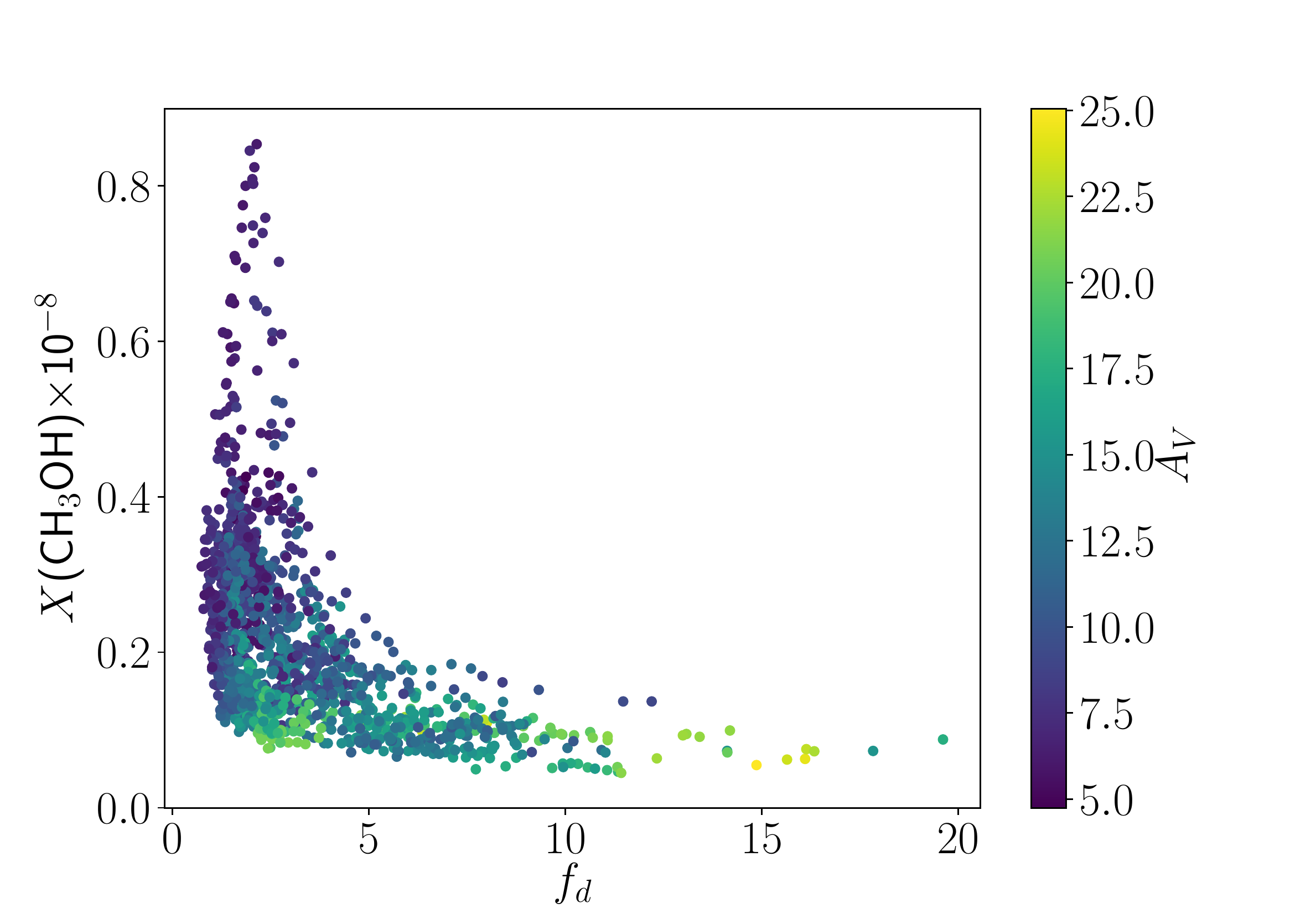}
\caption{The observed abundance of CH$_3$OH for all of the pixels in all of the maps combined as a function of observed abundance of C$^{18}$O (left) and CO-depletion factor (right). The color scale represents the line-of-sight visual extinction.} \label{fig:X-meth-vs-X-co}
\end{figure*}

We use the $N$(H$_2$) obtained by fitting spectral energy distribution (SED) to the {\it Herschel} data to measure the methanol abundance. Methanol abundance maps are shown in Fig.~\ref{fig:x_meth_maps}. We obtain the methanol abundances of (0.5--8.5)$\times 10^{-9}$ with a median value of 1.9$\times 10^{-9}$ (see median column densities, rotational temperature and abundance of methanol toward individual cores in Table~\ref{tab:core_age}). Cores 1, 6, 7, 35, which belong to the less evolved regions B10, B211, and B216, show higher methanol abundance than cores 10, 11, 16 from the more evolved B213 region, by a factor of two. The highest methanol abundance (8.5$\times10^{-9}$) is observed toward a local methanol peak in a shell around core~1. In all cores, the highest methanol abundance is observed in the shells around the cores, with $A_V\!=\!5-10$~mag (see the left panel of Fig.~\ref{fig:meth_vs_Av}) and $f_d=1.5-2.0$ (see the right panel of Fig.~\ref{fig:X-meth-vs-X-co}), and correlates with gaseous CO abundance (see the left panel of Fig.~\ref{fig:X-meth-vs-X-co}). CO depletion factor increases with visual extinction (see the right panel of Fig.~\ref{fig:meth_vs_Av}), however there is rather a general trend than a single correlation: $f_d$ always increases with $A_V$ within one core (see the plots of individual cores in Fig.~\ref{fig:compare_models}). Sometimes the emission has a prominent peak of enhanced abundance (cores 1 and 35), on average associated with $A_V\simeq6$~mag and $f_d\simeq1.6$, which is well illustrated by the abundance maps of cores 1 and 7 in Fig.~\ref{fig:x_meth_maps}. For cores 10, 11, and 16 we do not have enough data points to test if the average rule applies to them. In all cores methanol is significantly depleted toward the dust peaks where the observed abundance is minimal (except for core~35 where minimal abundance is observed to the north of the dust peak). 

In L1544 and Oph-H-MM1 the lopsided distribution of methanol in the shell was explained by the UV irradiation of one side and shielding of the other side of the cores \citep{Spezzano2016l,Harju2020} and the probable presence of a slow shock \citep{Punanova2018a,Harju2020}. In L1495, there are no preferred UV irradiation directions or areas shielded more than others \citep[the background column density on either sides of the filament differs by a factor of 2;][]{Palmeirim2013}. Although locally illumination of the cores could be non-uniform: there are protostars of various classes in and around the filament. Besides that, from \citet{Hacar2013} we know that the cores reside in the dense ``fertile'' fibers next to less dense ``sterile'' fibers, those  overlap along the line of sight and probably both shield methanol from UV destruction and contribute to the observed methanol abundance producing local abundance peaks like the ones toward cores 1 and 35. Yet there are no evidence of slow shocks in the L1495 cores: the local velocity gradients of the shells traced by the HCO$^+$ emission show homogeneous patterns \citep{Punanova2018b}, except for the protostellar core~11.  

\section{Modeling chemical composition of the filament cores}

\subsection{Physical models of the cores}\label{physmod}
\begin{figure*}\centering
\includegraphics[height=7.5cm,keepaspectratio]{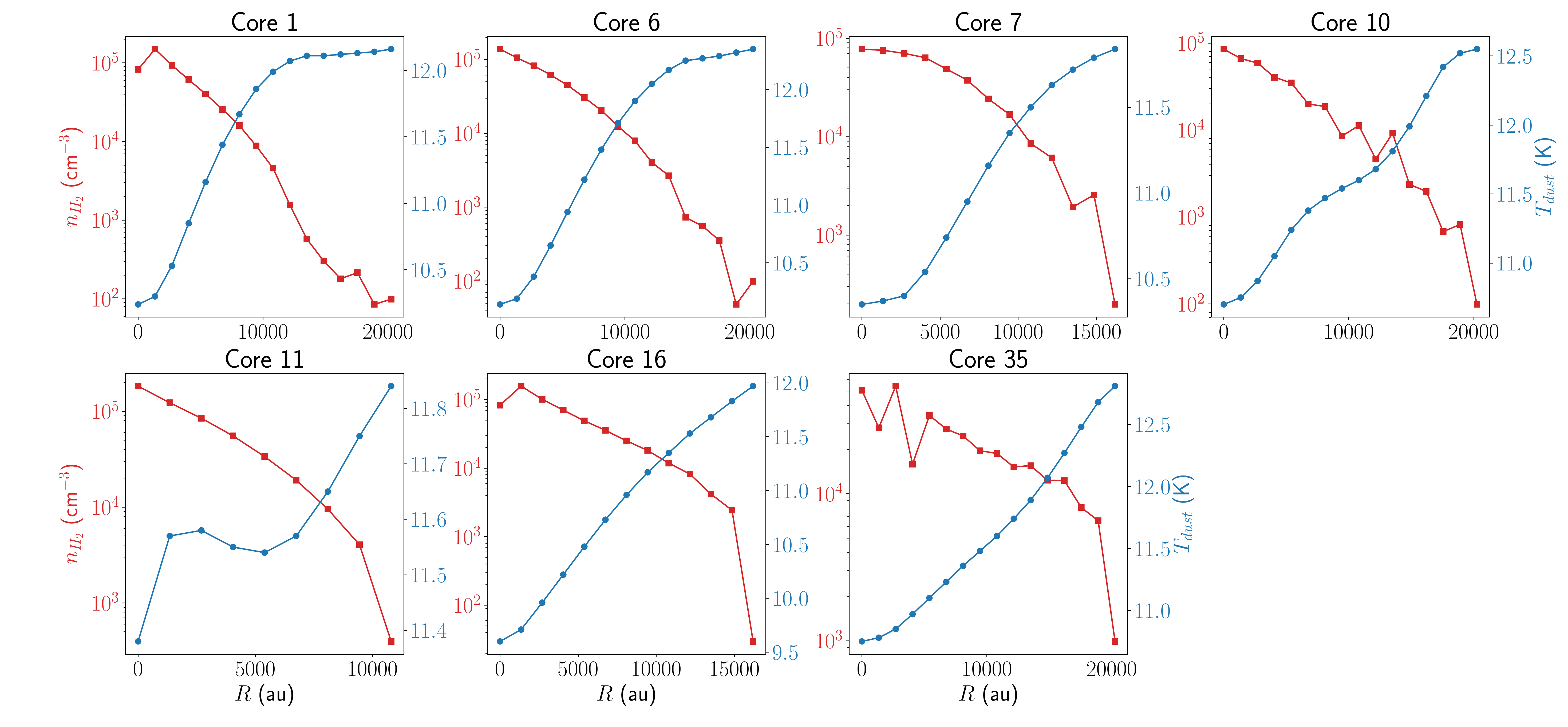}
\caption{Physical models of the cores: molecular hydrogen number density (red) and dust temperature (blue).}\label{fig:ave_model}
\end{figure*}

For the purpose of modeling, we assumed that the cores are spherically symmetric. Under this assumption, we derived temperature and density profiles of the cores. First, we used simple polynomial fits to the radial distributions of {\it Herschel} molecular column density ($N({\rm H_2})$). Then, we converted the molecular hydrogen column density to volume density. A similar approach was recently published in \citet[][]{HasenbergerAlves21}. Two key assumptions in the performed procedure are (i) the cores are spherical and (ii) the cores can be divided into constant gas density ``onion shells'' of same thickness. We also assume that the model center coincides with the observed dust peak. The derived H$_{2}$ volume densities have been tested to provide the original H$_{2}$ column densities when using the reverse procedure of obtaining column densities from volume densities described in \citet[][]{Jimenez-Serra2016}. The physical models of the cores are shown in Fig.~\ref{fig:ave_model}. Gas temperature in the cores was assumed equal to the dust temperature derived from {\it Herschel} data. This assumption is reasonable for our study, as we are focused on chemistry. The majority of chemical processes in the gas phase exhibit weak dependence on gas temperature. For example, rates of ion-molecule reactions that are believed to dominate low-temperature gas-phase chemistry, depend on temperature as $\sqrt{T_{\rm gas}}$. On the contrary, rates of chemical processes on surfaces of interstellar grains such as thermal desorption or thermal hopping of adsorbed species typically show exponential dependence on temperature. Moreover, at gas densities above 10$^{4}$~cm$^{-3}$ gas and dust are close to thermal coupling \citep{Goldsmith2001}. Finally, the L1495 filament does not possess signs of violent processes such as shock waves that may heat the gas to very high temperatures \citep[e.g.,][]{Tafalla2015}. Thus, we believe that our assumption on equality of gas and dust temperatures is justified.

\subsection{Chemical model}\label{sec:ch_model}

\begin{table}
    \caption{Median column densities, rotational temperatures and abundances of methanol for each core; the cores chemical ages, derived from the chemical model. The age estimates are based on the highest observed CO depletion factors.}
    \label{tab:core_age}
    \centering
    \begin{tabular}{cccccc}\hline\hline
        Core & $N$(CH$_3$OH) & $T_{\rm rot}$ & $X$(CH$_3$OH) & $f_d$ & age \\
         & (10$^{13}$~cm$^{-2}$) & (K) & $\times$10$^{-9}$ & & (kyr) \\
        \hline
        1 & 2.81 & 9.3 & 2.68 & 8 & 118 \\
        6 & 1.83 & 8.2 & 1.28 & 14 & 226 \\
        7 & 2.23 & 8.9 & 1.76 & 4 & 89  \\
        10 & 1.18 & 9.0 & 0.84 & 12 & 433 \\
        11 & 1.39 & 10.9 & 1.07 & 44 & 1000  \\ 
        16 & 1.70 & 8.6 & 0.88 & 17 & 248 \\
        35 & 2.24 & 9.2 & 1.89 & 4$^{a}$ & 187 \\
         \hline
    \end{tabular}
    \begin{flushleft} Notes. $^a$We have no CO data for core~35, so we adapt the lowest $f_d$ observed toward our cores ($f_d$=4 for core~7), since the region B216 is supposed to be young \citep{Hacar2013,Seo2015}. 
    \end{flushleft}
\end{table}

In this work, we utilize the MONACO chemical model previously described in \citet{Vasyunin2017} with several minor updates described in~\citet[][]{Nagy2019,Scibelli2021}. This is a ``three-phase'' chemical model capable of simulating chemical evolution in the gas phase, on surfaces of icy mantles of interstellar dust grains, and in the bulk of icy mantles. MONACO is a 0D rate equations-based time-dependent chemical model. In order to simulate chemistry in pre-stellar cores, the chemical code is wrapped into a 1D static physical models that includes radial distributions of the most important physical parameters which control chemistry in a pre-stellar core: gas density, gas and dust temperatures, and visual extinction (see Sect.~\ref{physmod}). With the model, time-dependent fractional abundances calculated with the MONACO code at each radial point of the 1D model, can be converted to column densities of species, thus allowing direct comparison with observational data.

Since dense cores considered in this work are located in a filament of molecular gas, we use a two-step approach to model their chemical evolution. On the first step, chemistry in a low-density translucent medium is simulated ($n_{\rm H}$~=~10$^{2}$~cm$^{-3}$, $T_{\rm gas}$~=~$T_{\rm dust}$~=~20~K, $A_{V}$~=~2~mag) over a period of 10$^{6}$~years. The initial chemical composition of the medium corresponds to elemental abundances 1 (EA1) from Table~1 in \citet{Wakelam2008}. On the second step, chemical evolution in the core is simulated using the abundances of chemical species at the final moment of time of step one as initial chemical composition.

Other key parameters of the chemical model are the following. The cosmic-ray ionization rate is taken equal to the standard value of 1.3$\times$10$^{-17}$~s$^{-1}$. The dust grain radius in the model is 10$^{-5}$~cm. Four upper layers of multilayered grain mantle are considered as belonging to surface in accordance with~\citet[][]{VasyuninHerbst13mc}. The sticking probability for all species is unity  \citep[e.g.,][and references therein]{Fraser2004,Chaabouni2012}. Surface species can be delivered to the gas phase via thermal desorption, photodesorption, cosmic ray-induced desorption~\citep[][]{HasegawaHerbst93} and reactive desorption (RD).

Photodesorption yield for CO was measured in a number of laboratory experiments. It was found to be high, temperature- and wavelength-dependent, of the order of 10$^{-3}$--10$^{-1}$ molecules per incident photon \citep[e.g.,][]{Oeberg2009,Fayolle2011,Chen2014,MunozCaro2016,Paardekooper2016}. We tested various CO photodesorption yields (0.001--0.1) applied to our four models and found that high CO photodesorption yield, 0.03 molecules per incident photon and higher, significantly affects process of CO depletion in the models of pre-stellar cores considered in this study. For two cores, observed CO depletion factors either never reached (core~11), or reached on timescales $>$10$^6$ years (core~10). Such timescale is much longer than typical estimates of ``chemical age'' of pre-stellar cores, which is very uncertain, but in most studies is well below $10^6$ years: 1--3$\times\!10^5$ \citep[e.g.,][]{Tafalla2004b,Walsh2009,Jimenez-Serra2016,Nagy2019,Lattanzi2020,Scibelli2021,JimenezSerra_ea21}, sometimes up to 3--7$\times\!10^5$~years \citep[e.g.,][]{Loison2020}. Thus, with certain simplification, in this study we employed constant photodesorption yield for CO equal to 10$^{-2}$ molecules per incident photon, following \citet{Fayolle2011}, where this value was obtained for UV~field in pre-stellar cores in accurate experiments with tunable synchrotron radiation. For methanol, following \citet{Bertin2016} and \citet{Cruz-Diaz2016}, we employed photodesorption yield equal to 10$^{-5}$ molecules per incident photon. This yield was also assumed for other species in the model.

The efficiency of reactive desorption, i.e., the probability of a product of an exothermic surface reaction to be ejected to the gas phase upon formation depends on the particular reaction, product and type of surface in a complex and still poorly studied \citep[see e.g.,][]{Garrod2007,Vasyunin2013,Minissale2016,Chuang_ea18}. Next, the rate of reactive desorption depends on the rate of related surface reaction. Rates of surface diffusive reactions of hydrogenation are mainly controlled by diffusion of atomic hydrogen. The rates of diffusion, in turn, are poorly known. Reactive desorption is believed to be the key process that delivers methanol, formed on cold grains during CO hydrogenation, to the gas phase~\citep[][]{Garrod_ea06}. Given that there are no known efficient gas-phase routes of methanol formation \citep[][]{Geppert_ea06}, it is tempting to utilize the extensive observational data set on CO and CH$_{3}$OH presented in this study, to constrain model parameters related to the formation and desorption of methanol in the conditions of cold dense cores. Thus, we consider four parametrizations of reactive desorption and surface mobility of hydrogen atoms: (i) tunneling for hydrogen diffusion enabled, treatment of reactive desorption following \citet[][]{Minissale2016} with parameters as in \citet[][]{Vasyunin2017} (rdMD); (ii) no tunneling for hydrogen diffusion, only thermal hopping, diffusion/desorption energy ratio is 0.5, rdMD; (iii) tunneling for hydrogen diffusion, single reactive desorption probability of 1\% (rd1); (iv) no tunneling for hydrogen diffusion, rd1. Those four particular models were chosen based on the fact that efficiency of quantum tunneling for atomic diffusion on grains is still debated  \citep[e.g.,][]{Rimola2014,Lamberts_2017}. As of parametrizations of reactive desorption, we aim to check the feasibility of approaches utilized in our previous works~\citep[e.g.,][]{Vasyunin2013, Vasyunin2017}, as well in a number of other studies \citep[e.g.,][]{Garrod2007,Ruaud2016,Jin2020}.

\begin{figure*}
    \centering
    \includegraphics[height=5.8cm,keepaspectratio]{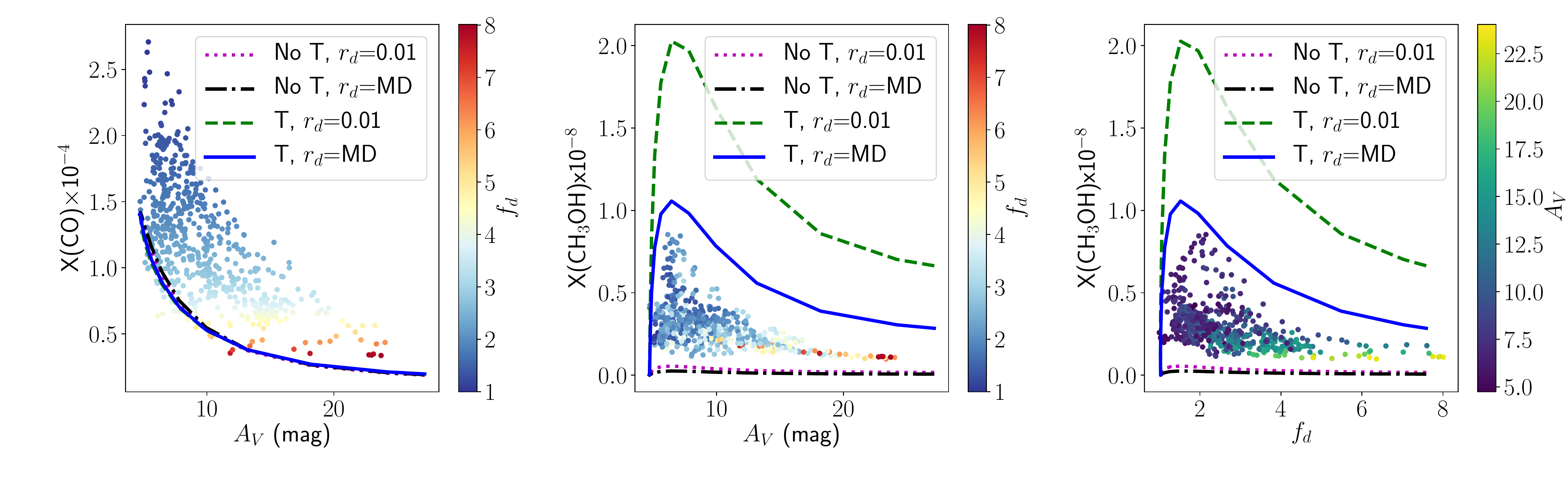}
    \caption{The comparison of the observed (colored dots) and modeled CO abundance profiles as a function of visual extinction (left); CH$_3$OH abundance profiles as a function of visual extinction (middle); and CO depletion factor (right) for core~1. The color scale represents CO depletion factor $f_d$ (left and middle) and visual extinction $A_V$ (right) in the individual pixels. The models include: no tunneling for diffusion of H and H$_2$ and 1\% reactive desorption efficiency (dotted magenta line); no tunneling and reactive desorption efficiency from \citet{Minissale2016} (dashed-dotted black line); tunneling and 1\% reactive desorption efficiency (dashed green line); tunneling and reactive desorption efficiency from \citet{Minissale2016} (solid blue line).}
    \label{fig:abundance_modelling_core1}
\end{figure*}

\begin{figure*}
	\centering
	\includegraphics[height=19cm,keepaspectratio]{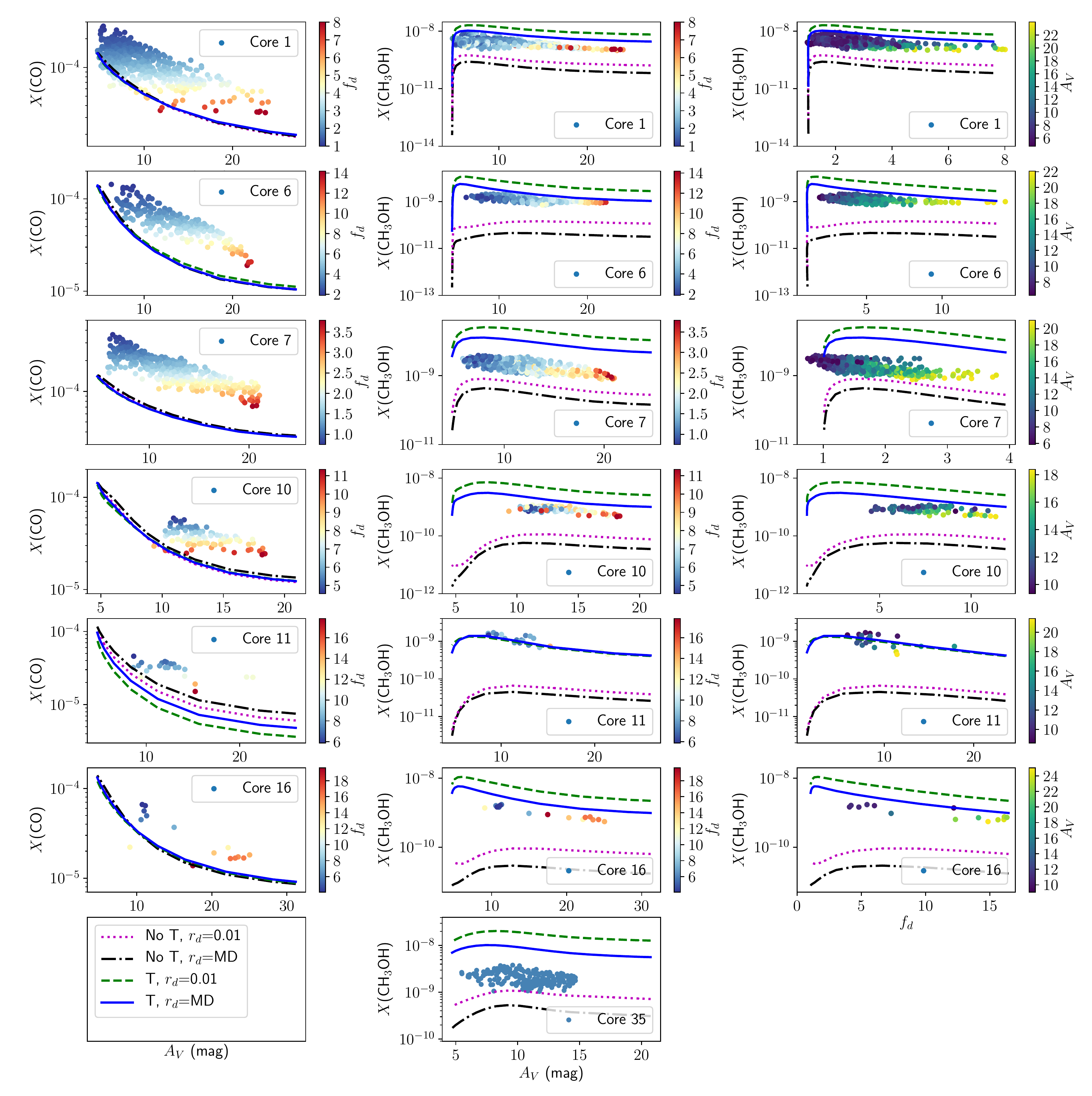}
	\caption{The comparison of the observed (colored dots) and modeled CO abundance profiles as a function of visual extinction (left); CH$_3$OH abundance profiles as a function of visual extinction (middle); and CO depletion factor (right) for all cores. The color scale represents CO depletion factor $f_d$ (left and middle) and visual extinction $A_V$ (right) in the individual pixels. The data points of core~35 are shown with plane color since there are no CO data for this core. The models include: no tunneling for diffusion of H and H$_2$ and 1\% reactive desorption efficiency (dotted magenta line); no tunneling and reactive desorption efficiency from \citet{Minissale2016} (dashed-dotted black line); tunneling and 1\% reactive desorption efficiency (dashed green line); tunneling and reactive desorption efficiency from \citet{Minissale2016} (solid blue line).}
	\label{fig:compare_models}
\end{figure*}

\subsection{Observed and modeled abundances}
Using the parameters described above, we performed 1D chemical modeling of all dense cores in the L1495 filament considered in this study. To estimate the chemical age of the cores, we used the highest depletion factor values reached in each core (see Table~\ref{tab:core_age}). Modeled radial profiles of CO and CH$_{3}$OH column densities and fractional abundances were then compared with the observed values. The four models produce almost the same CO abundance (see left panel of Fig.~\ref{fig:abundance_modelling_core1}), while methanol abundance differ significantly from model to model (see central and right panels of Fig.~\ref{fig:abundance_modelling_core1}). Although none of the models matches the observations perfectly, the model with quantum tunneling for diffusion of atomic hydrogen and treatment of reactive desorption according to parametrization by~\citet[][]{Minissale2016} produces results with the closest agreement to the observed data, as shown in Fig.~\ref{fig:abundance_modelling_core1} with an example of core~1. The left panels of Fig.~\ref{fig:compare_models} show the comparison of the observed CO abundances vs. visual extinction and the results of all four chemical models.

The best model reproduces the observed CO abundances very well with a slight (by a factor of a few) underestimation likely due to the fact that we see the CO emission from the cloud surrounding the cores. The other models (with higher reactive desorption and no tunneling) produce the same amount of CO (see left panel of Fig.~\ref{fig:abundance_modelling_core1} and~\ref{fig:compare_models}). This result shows that the cores are close to spherical shape, and our assumption used to convert gas column density to volume density is applicable for our cores. 

In the middle panels of Fig.~\ref{fig:compare_models}, observed and modeled abundances of methanol are presented. One can see that CH$_{3}$OH abundances obtained with the best fit model are overestimated by a factor of a few in all cores, similar to what was seen before in the models \citep[like in L1521E in][]{Scibelli2021}, except core~11. In that core, the model matches the observed values well. However, the number of data points for core~11 is small. Another reason for the mismatch might be our assumption of the $A$:$E$ methanol ratio of 1:1. If we treat the two forms separately as is shown in Sect.~\ref{sec:col_den}, $E$-methanol shows higher column densities than $A-$, although with large uncertainties. On the other hand, \citet{Harju2020} present the ratio $A$:$E$=1.3. The complex correlation between methanol abundance and visual extinction is clearly seen both in the model and observations, as illustrated in Fig.~\ref{fig:colden_av_good_model}: we detect the lowest methanol column densities at the outskirts of the cores, they rapidly increase with $A_V$ up to the line-of-sight $A_V\sim8$, and then decrease with $A_V$.

In the right panels of Fig.~\ref{fig:compare_models}, abundances of methanol vs. depletion factor $f_d$ are presented. Interestingly, the maximum abundances of CH$_{3}$OH correspond to moderate values of CO depletion factor, $f_d\!=\!1.5$--2, which is associated with volume densities of $\sim\!10^4$~cm$^{-3}$ and temperature of $\sim\!11.5$~K. This fact supports the scenario of formation of methanol during the onset of catastrophic CO freeze-out in pre-stellar cores \citep{Bizzocchi2014,Vasyunin2017}. The almost linear correlation between the CH$_3$OH and CO abundances shown in the left panel of Fig.~\ref{fig:X-meth-vs-X-co} might be related to the efficiency of CO hydrogenation to methanol on grains, as was suggested by \citet{Harju2020}. 
 
Figures~\ref{fig:abundance_modelling_core1} and~\ref{fig:compare_models} show that models without tunneling for hydrogen diffusion severely underproduce methanol abundances, while model with tunneling and single 1\% probability of reactive desorption overproduces the abundance of methanol by more than one order of magnitude in the majority of the cores. The only exception is core~7, where methanol abundance is better reproduced by the model without tunneling and 1\% probability of reactive desorption. This is the core with the lowest CO depletion factor ($f_d$=4) in this study. Low $f_d$ value can be attributed to the very young chemical age of the core \citep[e.g., L1521E,][]{Nagy2019,Scibelli2021}. For such a young core, our modeling approach based on static density profile and ``low metals'' elemental chemical composition is most likely less accurate than for other, more evolved cores considered in this study. To summarize, observational data on CO and CH$_{3}$OH in the dense cores of the L1495 filament can be fitted best with the MONACO code when tunneling for atomic hydrogen is enabled, and reactive desorption is treated following the approach by~\citet[][]{Minissale2016}.

\section{Discussion} \label{sec:discussion}

In this work, we present an extensive observational study of methanol distribution in dense cores within the L1495 filament. The wealth of observational data is analyzed statistically, and studied with sophisticated numerical gas-grain model of chemical evolution in star-forming regions. Modeling results, especially the spatial distribution of species, agree surprisingly well with observational data given the simplicity of the physical models adopted to represent dense cores in this study. The cores are approximated as static spherical objects, with density and temperature gradients. While such approximation is often considered too simplistic, it appears to be sufficiently accurate for modeling of distribution of chemical species in dense cores. This fact may suggest that dense pre-stellar cores do not significantly deviate from spherical symmetry and evolve dynamically on timescales longer than timescales of chemical evolution. Characteristic timescales of chemical evolution in cores of average volume density of $10^5$~cm$^{-3}$ are of the order of 10$^{5}$~years, close to free-fall times for starless cores with similar average volume densities \citep[see, e. g.,][]{Kirk2005,Tafalla2006,Andre2014}. However, even the most dynamically evolved pre-stellar core (L1544) has been found to contract at a much smaller rate than free-fall \citep{Keto2010}, probably due to the retarding action of magnetic fields. Therefore, chemistry indeed proceeds faster than dynamical evolution at core scales. 

The extensive set of observational data allowed us to put constraints on the treatment of reactive desorption, at least for the astrochemical model that includes only diffusive surface chemistry. The observed abundances of methanol in the gas phase of pre-stellar cores can be reproduced most accurately with the model that includes quantum tunneling for diffusion of atomic hydrogen on surfaces of grains and parametrization of probabilities of reactive desorption following~\citet[][]{Minissale2016}. Model with those assumptions is presented in~\citet[][]{Jimenez-Serra2016} and \citet[][]{Vasyunin2017}. The model successfully reproduced the abundances of many molecules including COMs in L1544 and several other cores~\citep[e.g.,][]{Nagy2019, Lattanzi2020, JimenezSerra_ea21}. On the other hand, in the case of a very young core L1521E the model overpredicted methanol abundance and underpredicted the abundances of COMs \citep{Scibelli2021}.

In~\citet[][]{Vasyunin2017}, the maximum abundance of methanol in L1544 pre-stellar core is located near local $A_V$~=~4~mag (that is $A_V\sim8$ on the line of sight). This value is similar for other cores. It corresponds to the location in spherical clouds where catastrophic freeze-out of CO molecules starts. The onset of such catastrophic CO freeze-out corresponds to the highest accretion rate of carbon monoxide onto grains, thus facilitating the most efficient methanol formation across the core. Analysis of observations presented in this study, confirm that methanol abundances reach their maxima in different cores at similar values of visual extinctions. Moreover, the locations of CH$_3$OH abundance maxima typically correspond to moderate values of CO depletion factor, $f_d\simeq 2$. Thus, earlier conclusions on methanol formation scenario based on limited observational data, are now confirmed on a more significant data set.

Although this study advocates for tunneling as a source of mobility of hydrogen atoms and molecules on grains, one has to bear in mind that the model utilized in this work includes only diffusive surface chemistry. Tunneling increases the pace of CO hydrogenation into formaldehyde and methanol. However, the conclusion may change if chemical models that include non-diffusive grain chemistry will be applied to the presented data. One can speculate about very fast non-diffusive mechanisms of formation of those two species which may render diffusive tunneling not so important \citep[such as recombination of various CH$_{\rm n}$O radicals  produced in close proximity to each other on the icy grain surface,][]{Fedoseev2015,Jin2020}. On the other hand, one shall bear in mind that CO hydrogenation sequence may include efficient H$_2$ abstraction reactions. Those reactions may reduce the rate of H$_2$CO and CH$_3$OH formation. In any case, statistical data on CO and CH$_3$OH abundances and their spatial distribution in cold cores of the L1495 filament is a unique tool to test current and future gas-grain chemical models.

Chemistry in dense cores and clumps is normally modeled with the so-called low-metal abundances \citep[e.g.,][]{Hasegawa1992,Lee1998,Sipila2020} summarized in \citet{Wakelam2008} following \citet{Graedel1982}. While the low- and high-metal abundances are supposed to represent the abundances of such heavy elements as Fe, Mg, Si, S, etc., CNO elements also have different abundances in these two sets. \citet{Graedel1982} note that poorly constrained metal abundances correlate with (and influence) electron fraction; high-metal abundance corresponds to $x(e^-)>$few$\times$10$^{-7}$ at $n\simeq10^4-10^5$~cm$^{-3}$. They also note that the relative abundances change with age, in particular, between 10$^5$ and 10$^7$~yr (the typical chemical age of a pre-stellar core is 1--3$\times$10$^5$) the abundances may change by an order of magnitude. \citet{Caselli1998} showed that the electron fraction depends on carbon and oxygen elemental depletion; they find $x(e^-)\simeq2.5\times10^{-7}$ for L1495 (with assumption of high elemental depletion), which is similar to our estimation $x(e^-)\sim2\times10^{-7}$ based on DCO$^+$ and H$^{13}$CO$^+$ observations (Punanova et al. in prep.). Another argument against standard ``low metals'' elemental abundances appeared in the course of this study. The best-fit reference undepleted abundance of CO for L1495 confirmed by this work is 2.7$\times$10$^{-4}$ w.r.t. H$_2$. This implies elemental abundances of both carbon and oxygen available for gas phase chemistry to be at least as high as that value. At the same time, ``low metals'' values for C and O abundances are 1.46$\times$10$^{-4}$ and 3.52$\times$10$^{-4}$ w.r.t. H$_2$, correspondingly~\citep[][]{Wakelam2008}, while ``high metals'' elemental abundances of C and O listed in that paper are 2.4$\times$10$^{-4}$ and 5.12$\times$10$^{-4}$ w.r.t. H$_2$. These considerations allow us at least to put in question the use of the low-metal abundances and suppose that probably high-metal abundances are more suitable for modeling dense cores in Taurus, as long as freeze-out is taken into account. Note however that direct application of ``high metals'' initial abundances leads to overestimation of abundances of sulfur-bearing species in models \citep{Shalabiea2001}, unless S is heavily depleted onto dust grains \citep{Caselli1994}. Besides, \citet{Scibelli2021} showed that elevated sulfur abundance, needed to reproduce the abundances of S-bearing species \citep[like it was also done by][]{Seo2019}, leads to decrease in the abundances of COMs (e.g., CH$_3$OCH$_3$). On the other hand, the value of elemental sulfur abundance may deserve additional considerations, since understanding of sulfur chemistry is currently actively developing and still far from maturity~\citep[][Cazaux et al. sbmt.]{Nagy2019, LaasCaselli19, Shingledecker_ea20}.

\section{Conclusions}\label{sec:conclusions}

In this paper we study spatial distribution of methanol in cold dense cores. We explore correlations between the methanol abundance and visual extinction and CO depletion. We test the three-phase chemical model MONACO \citep{Vasyunin2017} against a large and homogeneous data set of 3~and 2~mm maps of methanol emission toward seven cold dense cores embedded in the L1495 filament. We vary key chemical model parameters to find the best match between the model and observational results. Our main findings are presented below.

\begin{enumerate}
    \item The highest methanol intensity is observed both toward shells around the dust peaks of the cores (1, 7, 10) and toward the dust peaks (cores 6, 11, 16, 35). The column densities vary from 0.5 to 7$\times10^{13}$~cm$^{-2}$.
    \item The highest methanol gas abundance is observed in the shells around the cores (this is also true for those cores that have the CH$_3$OH peak located at the dust peak), while toward the dust peaks methanol is depleted. We obtain the methanol abundances of (0.5--8.5)$\times$10$^{-9}$ with a median value of 1.9$\times$10$^{-9}$. 
    \item We analyze CO gas abundance in the entire filament based on the C$^{18}$O data from \citet{Tafalla2015} and conclude that the most suitable total undepleted CO abundance for the L1495 filament is 2.7$\times$10$^{-4}$ \citep{Lacy1994}. We find CO depletion factor within the cores from 1 to 44, with highest depletion factor toward the dust peaks from 4 to 44.
    \item As expected, methanol abundance increases with visual extinction at low $A_V$, and reaches maximum values around line-of-sight $A_V\!\simeq5$--8~mag and decreases at higher $A_V$.
    \item The highest methanol abundance is observed around moderate values of $f_d\!\simeq$1.5--2. This fact favors the scenario of methanol formation during the catastrophic freeze-out stage in starless cores.
    \item Dense cores take various shapes; even in the plane of sky the cores are far from being circular. However, for simplicity, we use radial temperature and volume density gradients as base for the chemical model, and the modeled abundances reproduce well the observed ones, which implies that spherically symmetric models overall provide a good match to single-dish observations of dense cores.
    \item We find that H and H$_2$ surface diffusion via tunneling is essential to reproduce the observed abundances of methanol. The model with disabled tunneling underproduces methanol abundance by 1.5 orders of magnitude.
    \item We compared the models with flat 1\% effectiveness of reactive desorption and the one calculated following the empiric formula yielded by \citet{Minissale2016}. The model best reproduces methanol abundance with the \citet{Minissale2016} reactive desorption effectiveness, while flat 1\% effectiveness tends to overproduce methanol abundance by a factor of a few. 
    \item Our observation results could serve as a benchmark for the forthcoming chemical models.
\end{enumerate}

\begin{acknowledgments}
The authors thank the anonymous referee for valuable comments that helped to improve the manuscript. A.V. and A.P. are the members of the Max Planck Partner Group at the Ural Federal University. A.P. and A.V. acknowledge the support of the Russian Science Foundation project 18-12-00351 and of the Russian Ministry of Science and Education via the State Assignment Contract no. FEUZ-2020-0038 (discussion on elemental abundances). The authors thank Vadim Krushinskiy for his help with Matplotlib and Gleb Fedoseev for discussion about photodesorption yield.
\end{acknowledgments}

%

\vspace{5mm}
\facilities{IRAM~30m, {\it Herschel}}


\software{astropy \citep{Astropy2013,Astropy2018},  
          }



\appendix

\section{Column density calculation}\label{app:col_den}

\begin{figure}
    \centering
    \includegraphics[height=6.0cm,keepaspectratio]{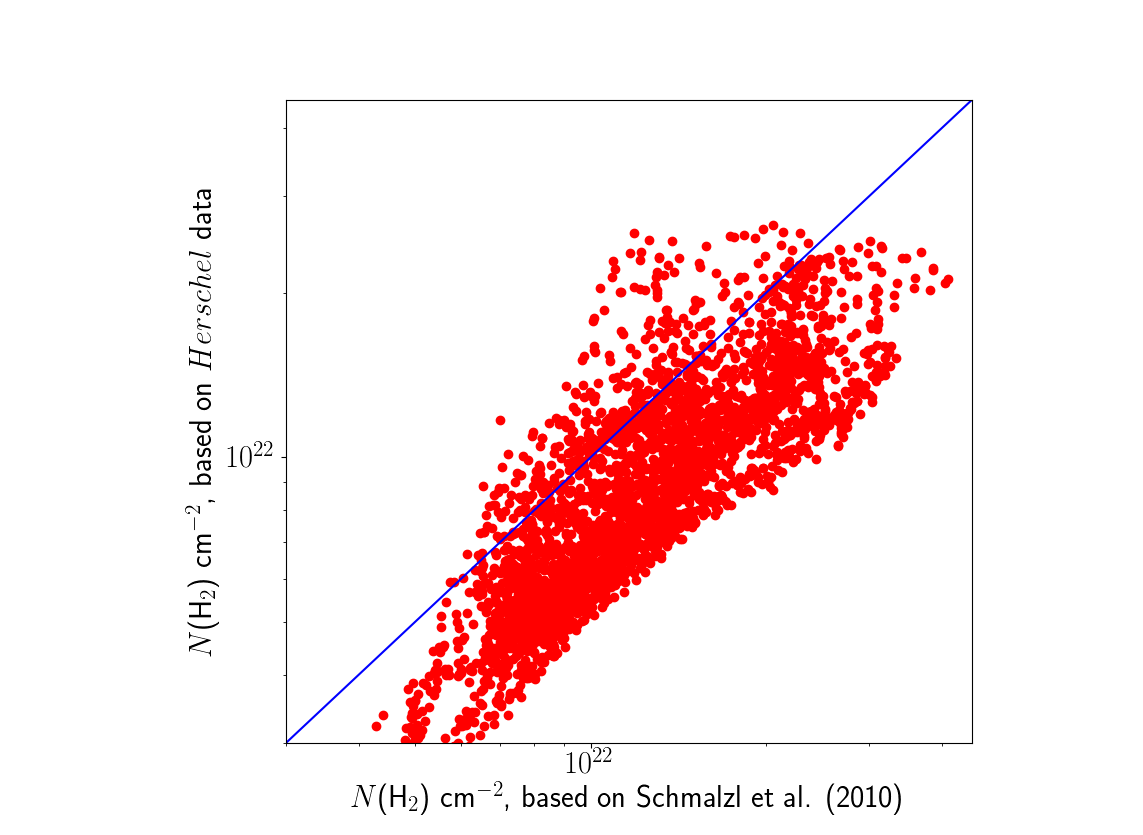}
    \caption{A comparison of $N({\rm H_2})$ measured via {\it Herschel} dust continuum emission and via $A_V$ measurements from infrared photometric observations \citep[from][]{Schmalzl2010}. The solid blue line corresponds to 1:1 correlation.}
    \label{fig:herschel-vs-av}
\end{figure}

We measure the methanol column densities via the rotational diagrams based on the four brightest lines detected across all cores (see Sect.~\ref{sec:col_den}). We use only the pixels with all four lines detected with signal-to-noise ratio $S/N\!>\!5$. We assume the methanol lines are optically thin \citep[as is shown in][]{Scibelli2020}, consistent with LTE, and the fractional abundance of $A-$ and $E-$methanol is $1\!:\!1$. We note that $A\!:\!E$ methanol ratio may differ from 1 \citep[1.2--1.5 in the starless core H-MM1 in Ophiuchus and 1.3  in this filament, L1495, from RADEX modelling,][]{Harju2017,Scibelli2020}, while we use two $E$-lines and two $A$-lines for our rotational diagrams. To test the 1:1 ratio assumption, we took the data from the brightest pixels, where all six methanol lines, observed in the project, were well detected.

\begin{figure}
    \centering
    \includegraphics[height=5.5cm,keepaspectratio]{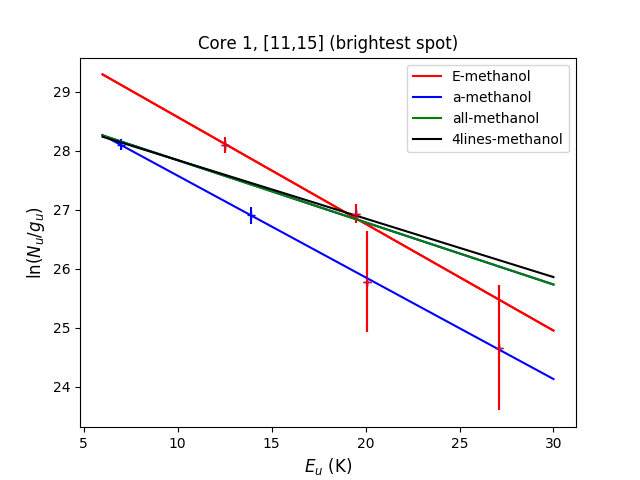}
    \caption{A comparison of methanol rotational diagrams toward the brightest pixel in core~1: based on E-methanol only (red line), based on A-methanol only (blue), based on all six A- and E-methanol lines including the weak higher energy lines (green), and based on the four A- and E-methanol lines (black, the one used to measure $N_{\rm tot}$). Red symbols show E-methanol, blue symbols show A-methanol.}
    \label{fig:rot_diag_comparison}
\end{figure}

Figure~\ref{fig:rot_diag_comparison} shows the rotational diagrams plotted for one position with the brightest methanol emission in core~1, based on two $A$-methanol lines (blue), four $E$-methanol lines (red), four brightest methanol lines (used in the paper, black), and all six methanol lines (green). The resulting column densities and rotational temperatures are presented in Table~\ref{tab:rot_diag}. When we consider $A$- and $E$-methanol separately, the rotational diagrams give lower rotational temperature, similar for both forms, 5.5--5.8~K (steeper slope of the diagrams), than that from the common rotational diagrams (9.5--10.1~K). 

The total column density of $A$-methanol is lower than that of $E$-methanol, although the difference is within their large uncertainties. Previous works present prevalence of $A$-methanol over $E$-methanol \citep[$A\!:\!E$=1.3 in][]{Harju2020,Scibelli2020}. The column densities based on the $A\!+\!E$ rotational diagrams are in agreement with those measured by \citet{Scibelli2020} with RADEX. The assumption of the $A\!:\!E$=1.3 ratio would give even higher total column densities than those presented in this work. Given the small number of the observed lines and the large uncertainties of the individual rotational diagrams, we decided to use both $A$- and $E$-methanol for the rotational diagrams. The difference between the results of the rotational diagrams based on four and six lines is negligible when error bars are taken into account, so we can rely on the four-points-based diagrams which we can plot in the majority of the pixels in our maps.

\begin{table}
    \caption{The results of the rotational diagrams based on different sets of the methanol lines.}
    \label{tab:rot_diag}
    \centering
    \begin{tabular}{lcccc}
    \hline\hline
    & CH$_3$OH-$A$ & CH$_3$OH-$E$ & $A$ + $E$ 4 lines & $A$ + $E$ 6 lines \\
    \hline
     $N_{\rm tot}$ (10$^{13}$~cm$^{-2}$)& 4.6$\pm$1.5 & 12.5$\pm$6.3 & 6.6$\pm$1.5 & 6.4$\pm$1.3 \\
   $T_{\rm rot}$ (K) & 5.8$\pm$0.9 & 5.5$\pm$0.9 & 10.1$\pm$1.4 & 9.5$\pm$1.2 \\
    \hline
    \end{tabular}
\end{table}

\section{Additional figures}

\begin{figure*}
    \centering
    \includegraphics[height=4.2cm,keepaspectratio]{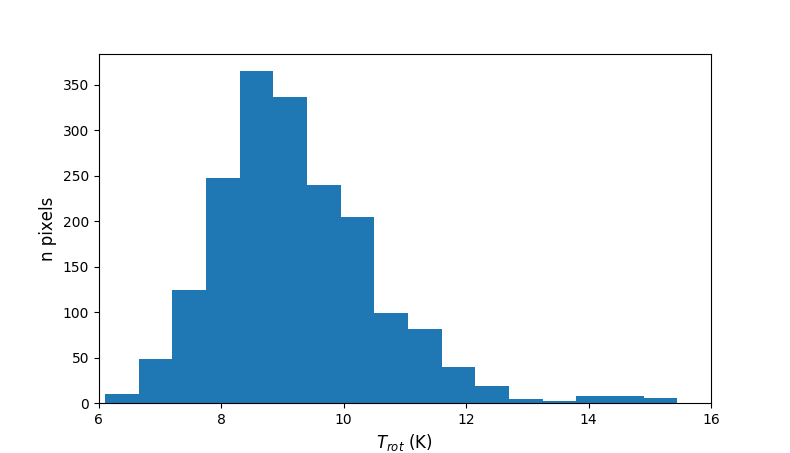}
    \includegraphics[height=4.2cm,keepaspectratio]{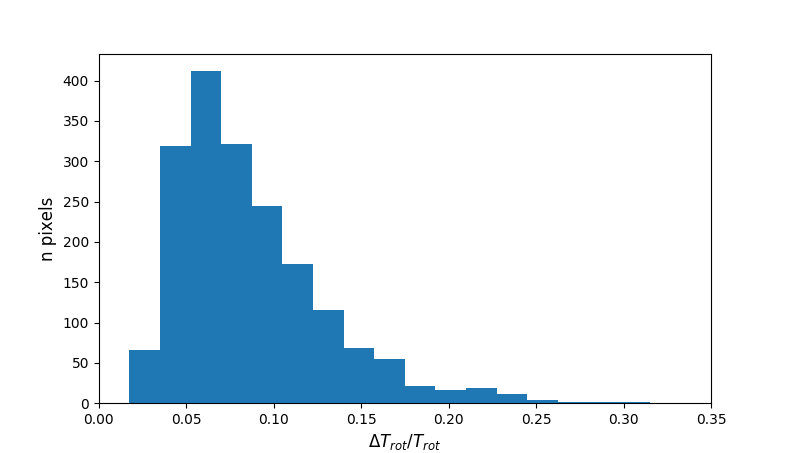}
    \caption{Distribution of $T_{\rm rot}$ based on the rotational diagrams (left) and the distribution of the relative uncertainties $\Delta T_{\rm rot}/T_{\rm rot}$ based on the rotational diagrams (right) of all available points with $S/N\!>\!5$ in the four lines used for the rotational diagrams.}
    \label{fig:t_rot}
\end{figure*}

\begin{figure*}
    \centering
    \includegraphics[height=4.2cm,keepaspectratio]{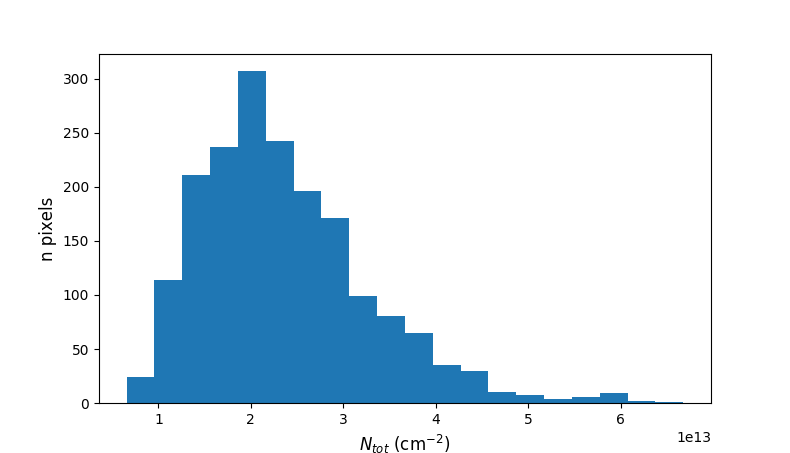}
    \includegraphics[height=4.2cm,keepaspectratio]{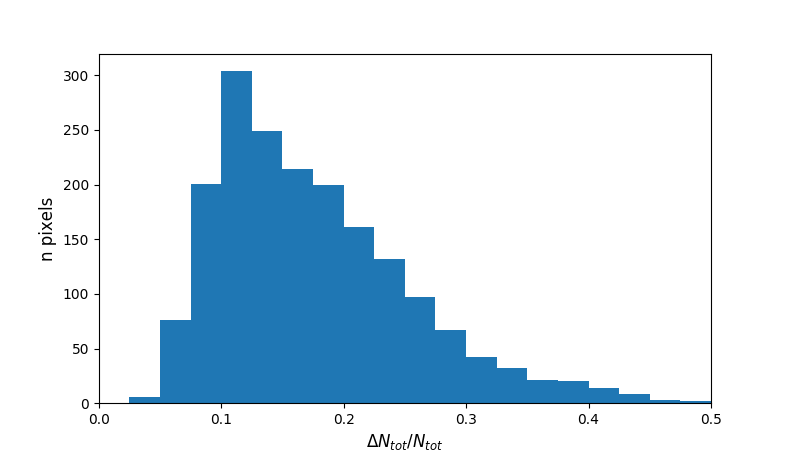}
    \caption{Distribution of $N_{\rm tot}$ based on the rotational diagrams (left) and the distribution of the relative uncertainties $\Delta N_{\rm tot}/N_{\rm tot}$ based on the rotational diagrams (right) of all available points with $S/N\!>\!5$ in the four lines used for the rotational diagrams.}
    \label{fig:n_tot}
\end{figure*}

\begin{figure*}
    \centering
    \includegraphics[height=9.0cm,keepaspectratio]{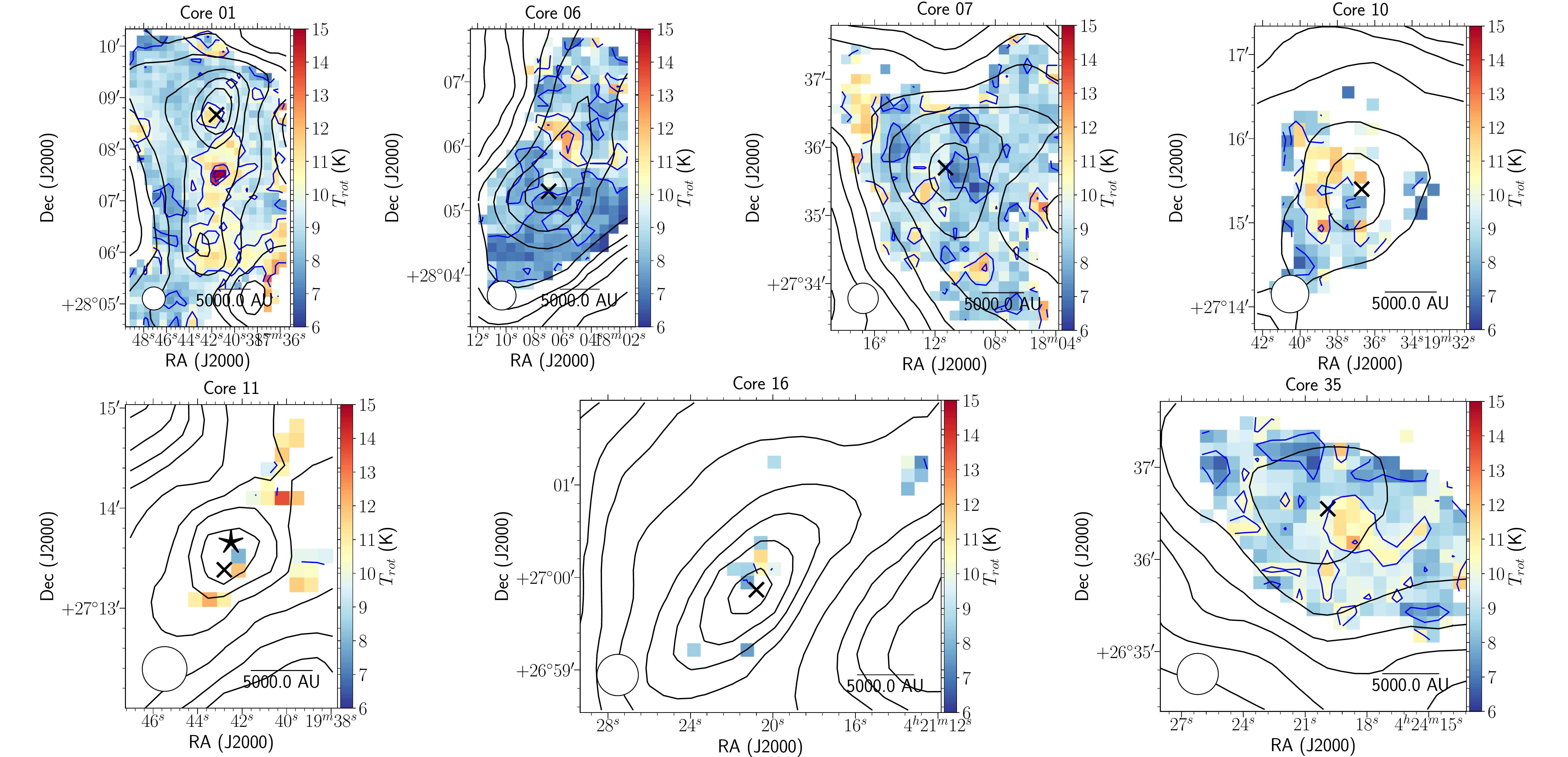}
    \caption{Rotational temperature line toward the observed cores (color scale and blue contours) and visual extinction (black contours at $A_V=$3, 4, 5, 6, 8, 12, 16, 20, 24~mag). The top $A_V$ contours are at $A_V$=24~mag for core~16 and at $A_V$=20~mag for the other cores. Black star shows the position of Class 0 protostar IRAS~04166+2706 \citep{Santiago-Garcia2009}, crosses show the {\it Herschel}~/~SPIRE dust emission peaks. White circle on the bottom left of each map shows the 26$^{\prime\prime}$ IRAM beam.}
    \label{fig:t_rot_maps}
\end{figure*}

\begin{figure*}
    \centering
    \includegraphics[height=9.0cm,keepaspectratio]{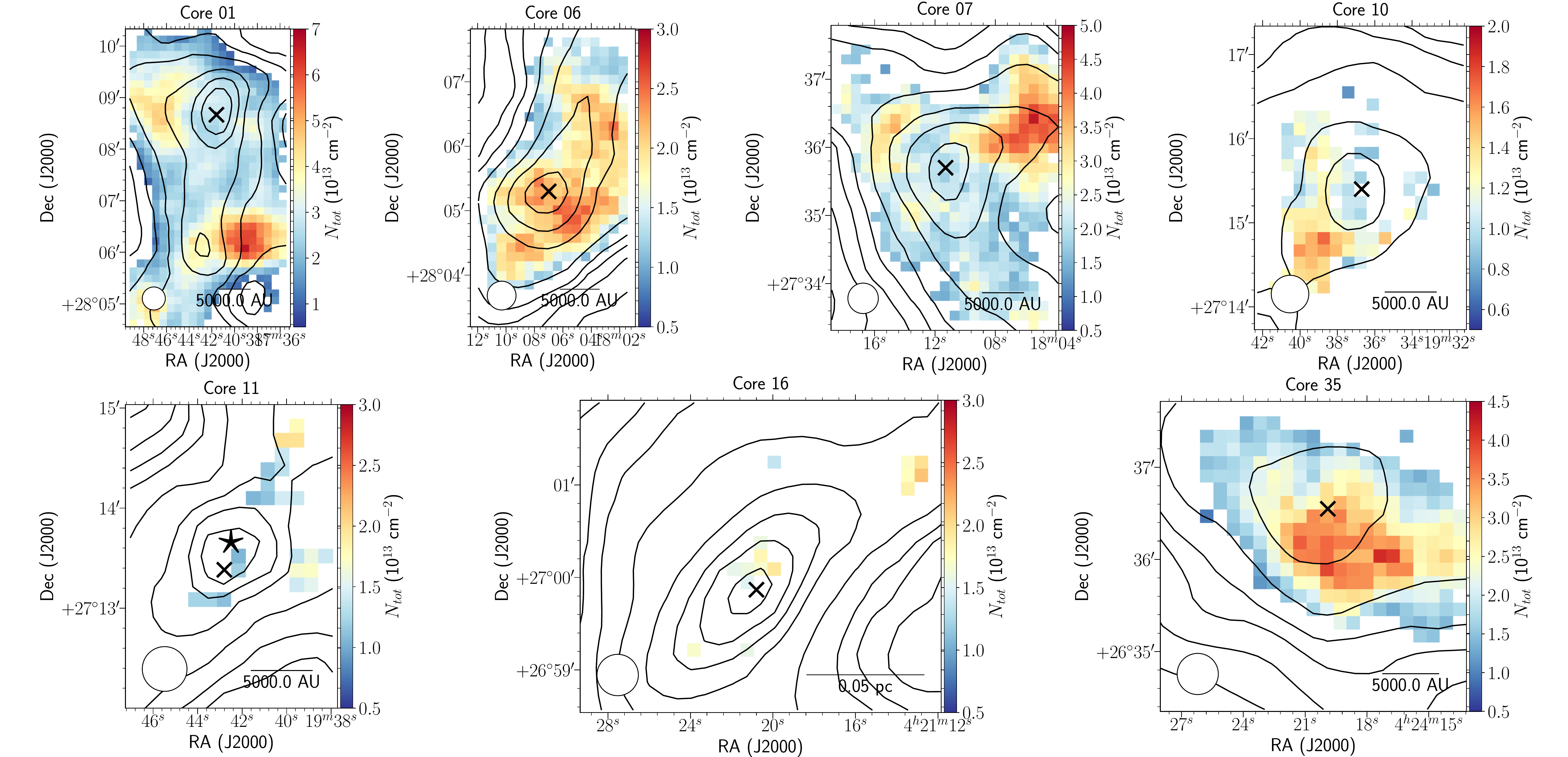}
    \caption{Methanol column densities toward the observed cores (color scale) and visual extinction (black contours at $A_V=$3, 4, 5, 6, 8, 12, 16, 20, 24~mag). The top $A_V$ contours are at $A_V$=24~mag for core~16 and at $A_V$=20~mag for the other cores. Black star shows the position of Class 0 protostar IRAS~04166+2706 \citep{Santiago-Garcia2009}, crosses show the {\it Herschel}~/~SPIRE dust emission peaks. White circle on the bottom left of each map shows the 26$^{\prime\prime}$ IRAM beam.}
    \label{fig:n_tot_maps}
\end{figure*}

\begin{figure*}
    \centering
    \includegraphics[height=9cm,keepaspectratio]{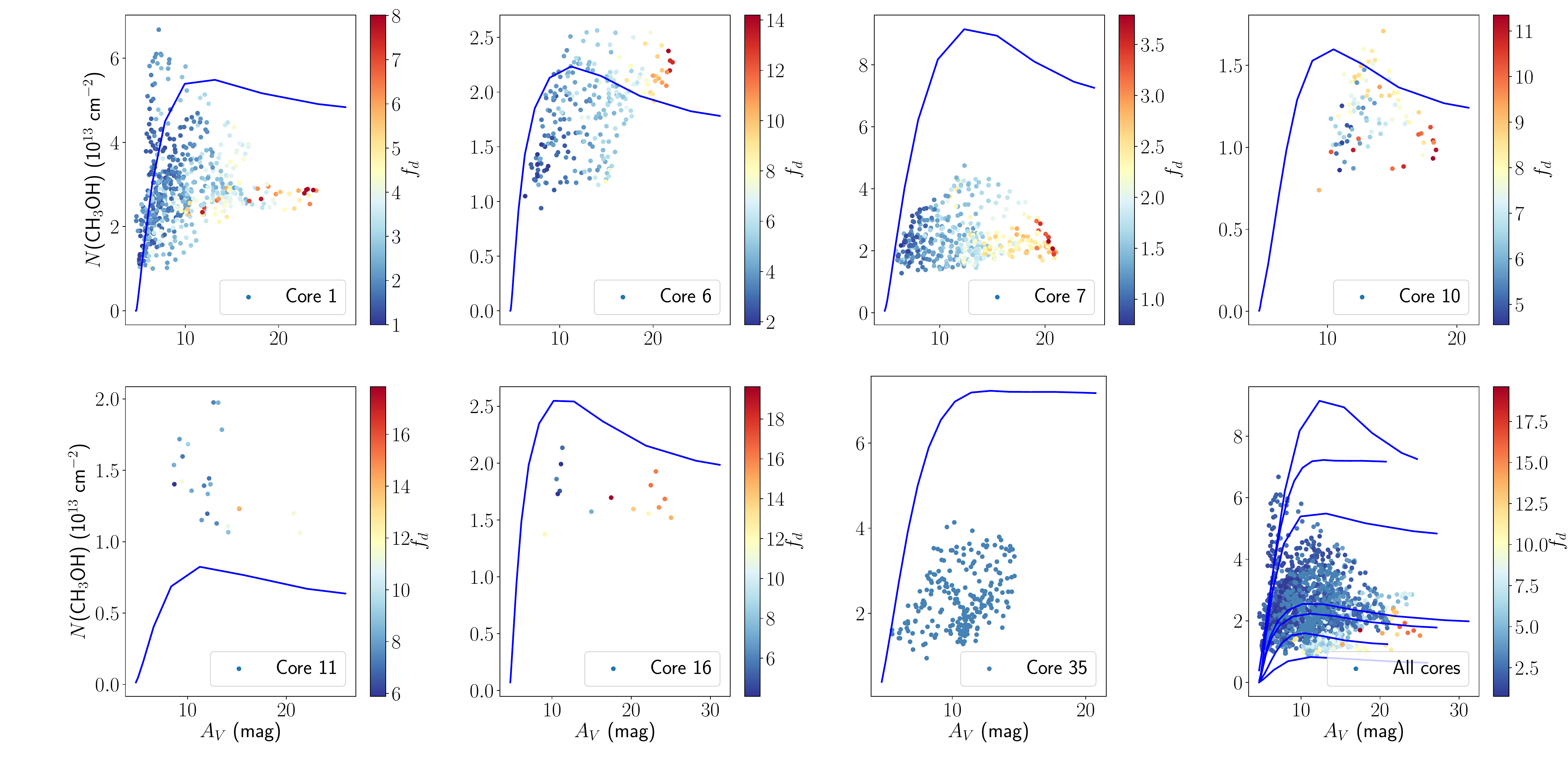}
    \caption{The comparison of the observed (colored dots) and modeled (solid blue line) CH$_3$OH column density profiles as a function of visual extinction. The color scale represents CO depletion factor $f_d$. The model includes tunneling for diffusion of H and H$_2$, and reactive desorption efficiency from \citet{Minissale2016}.}
    \label{fig:colden_av_good_model}
\end{figure*}


\bibliography{Punanova_lit}{}

\begin{thebibliography}{}
\expandafter\ifx\csname natexlab\endcsname\relax\def\natexlab#1{#1}\fi
\providecommand{\url}[1]{\href{#1}{#1}}
\providecommand{\dodoi}[1]{doi:~\href{http://doi.org/#1}{\nolinkurl{#1}}}
\providecommand{\doeprint}[1]{\href{http://ascl.net/#1}{\nolinkurl{http://ascl.net/#1}}}
\providecommand{\doarXiv}[1]{\href{https://arxiv.org/abs/#1}{\nolinkurl{https://arxiv.org/abs/#1}}}

\bibitem[{{Alonso-Albi} {et~al.}(2010){Alonso-Albi}, {Fuente}, {Crimier},
  {Caselli}, {Ceccarelli}, {Johnstone}, {Planesas}, {Rizzo}, {Wyrowski},
  {Tafalla}, {Lefloch}, {Maret}, \& {Dominik}}]{Alonso-Albi2010}
{Alonso-Albi}, T., {Fuente}, A., {Crimier}, N., {et~al.} 2010, \aap, 518, A52,
  \dodoi{10.1051/0004-6361/201014317}

\bibitem[{{Andr{\'e}} {et~al.}(2014){Andr{\'e}}, {Di Francesco},
  {Ward-Thompson}, {Inutsuka}, {Pudritz}, \& {Pineda}}]{Andre2014}
{Andr{\'e}}, P., {Di Francesco}, J., {Ward-Thompson}, D., {et~al.} 2014,
  Protostars and Planets VI, 27,
  \dodoi{10.2458/azu_uapress_9780816531240-ch002}

\bibitem[{{Astropy Collaboration} {et~al.}(2013){Astropy Collaboration},
  {Robitaille}, {Tollerud}, {Greenfield}, {Droettboom}, {Bray}, {Aldcroft},
  {Davis}, {Ginsburg}, {Price-Whelan}, {Kerzendorf}, {Conley}, {Crighton},
  {Barbary}, {Muna}, {Ferguson}, {Grollier}, {Parikh}, {Nair}, {Unther},
  {Deil}, {Woillez}, {Conseil}, {Kramer}, {Turner}, {Singer}, {Fox}, {Weaver},
  {Zabalza}, {Edwards}, {Azalee Bostroem}, {Burke}, {Casey}, {Crawford},
  {Dencheva}, {Ely}, {Jenness}, {Labrie}, {Lim}, {Pierfederici}, {Pontzen},
  {Ptak}, {Refsdal}, {Servillat}, \& {Streicher}}]{Astropy2013}
{Astropy Collaboration}, {Robitaille}, T.~P., {Tollerud}, E.~J., {et~al.} 2013,
  \aap, 558, A33, \dodoi{10.1051/0004-6361/201322068}

\bibitem[{{Astropy Collaboration} {et~al.}(2018){Astropy Collaboration},
  {Price-Whelan}, {Sip{\H{o}}cz}, {G{\"u}nther}, {Lim}, {Crawford}, {Conseil},
  {Shupe}, {Craig}, {Dencheva}, {Ginsburg}, {VanderPlas}, {Bradley},
  {P{\'e}rez-Su{\'a}rez}, {de Val-Borro}, {Aldcroft}, {Cruz}, {Robitaille},
  {Tollerud}, {Ardelean}, {Babej}, {Bach}, {Bachetti}, {Bakanov}, {Bamford},
  {Barentsen}, {Barmby}, {Baumbach}, {Berry}, {Biscani}, {Boquien}, {Bostroem},
  {Bouma}, {Brammer}, {Bray}, {Breytenbach}, {Buddelmeijer}, {Burke},
  {Calderone}, {Cano Rodr{\'\i}guez}, {Cara}, {Cardoso}, {Cheedella}, {Copin},
  {Corrales}, {Crichton}, {D'Avella}, {Deil}, {Depagne}, {Dietrich}, {Donath},
  {Droettboom}, {Earl}, {Erben}, {Fabbro}, {Ferreira}, {Finethy}, {Fox},
  {Garrison}, {Gibbons}, {Goldstein}, {Gommers}, {Greco}, {Greenfield},
  {Groener}, {Grollier}, {Hagen}, {Hirst}, {Homeier}, {Horton}, {Hosseinzadeh},
  {Hu}, {Hunkeler}, {Ivezi{\'c}}, {Jain}, {Jenness}, {Kanarek}, {Kendrew},
  {Kern}, {Kerzendorf}, {Khvalko}, {King}, {Kirkby}, {Kulkarni}, {Kumar},
  {Lee}, {Lenz}, {Littlefair}, {Ma}, {Macleod}, {Mastropietro}, {McCully},
  {Montagnac}, {Morris}, {Mueller}, {Mumford}, {Muna}, {Murphy}, {Nelson},
  {Nguyen}, {Ninan}, {N{\"o}the}, {Ogaz}, {Oh}, {Parejko}, {Parley}, {Pascual},
  {Patil}, {Patil}, {Plunkett}, {Prochaska}, {Rastogi}, {Reddy Janga},
  {Sabater}, {Sakurikar}, {Seifert}, {Sherbert}, {Sherwood-Taylor}, {Shih},
  {Sick}, {Silbiger}, {Singanamalla}, {Singer}, {Sladen}, {Sooley},
  {Sornarajah}, {Streicher}, {Teuben}, {Thomas}, {Tremblay}, {Turner},
  {Terr{\'o}n}, {van Kerkwijk}, {de la Vega}, {Watkins}, {Weaver}, {Whitmore},
  {Woillez}, {Zabalza}, \& {Astropy Contributors}}]{Astropy2018}
{Astropy Collaboration}, {Price-Whelan}, A.~M., {Sip{\H{o}}cz}, B.~M., {et~al.}
  2018, \aj, 156, 123, \dodoi{10.3847/1538-3881/aabc4f}

\bibitem[{{Bacmann} {et~al.}(2002){Bacmann}, {Lefloch}, {Ceccarelli},
  {Castets}, {Steinacker}, \& {Loinard}}]{Bacmann2002}
{Bacmann}, A., {Lefloch}, B., {Ceccarelli}, C., {et~al.} 2002, \aap, 389, L6,
  \dodoi{10.1051/0004-6361:20020652}

\bibitem[{{Barnard}(1927)}]{Barnard1927}
{Barnard}, E.~E. 1927, {Catalogue of 349 dark objects in the sky}

\bibitem[{{Beckwith} {et~al.}(1990){Beckwith}, {Sargent}, {Chini}, \&
  {Guesten}}]{Beckwith1990}
{Beckwith}, S. V.~W., {Sargent}, A.~I., {Chini}, R.~S., \& {Guesten}, R. 1990,
  \aj, 99, 924, \dodoi{10.1086/115385}

\bibitem[{{Benson} \& {Myers}(1989)}]{Benson1989}
{Benson}, P.~J., \& {Myers}, P.~C. 1989, \apjs, 71, 89, \dodoi{10.1086/191365}

\bibitem[{{Bergin} \& {Tafalla}(2007)}]{Bergin2007}
{Bergin}, E.~A., \& {Tafalla}, M. 2007, \araa, 45, 339,
  \dodoi{10.1146/annurev.astro.45.071206.100404}

\bibitem[{{Bertin} {et~al.}(2016){Bertin}, {Romanzin}, {Doronin}, {Philippe},
  {Jeseck}, {Ligterink}, {Linnartz}, {Michaut}, \& {Fillion}}]{Bertin2016}
{Bertin}, M., {Romanzin}, C., {Doronin}, M., {et~al.} 2016, \apjl, 817, L12,
  \dodoi{10.3847/2041-8205/817/2/L12}

\bibitem[{{Bizzocchi} {et~al.}(2014){Bizzocchi}, {Caselli}, {Spezzano}, \&
  {Leonardo}}]{Bizzocchi2014}
{Bizzocchi}, L., {Caselli}, P., {Spezzano}, S., \& {Leonardo}, E. 2014, \aap,
  569, A27, \dodoi{10.1051/0004-6361/201423858}

\bibitem[{{Bracco} {et~al.}(2017){Bracco}, {Palmeirim}, {Andr{\'e}}, {Adam},
  {Ade}, {Bacmann}, {Beelen}, {Beno{\^\i}t}, {Bideaud}, {Billot}, {Bourrion},
  {Calvo}, {Catalano}, {Coiffard}, {Comis}, {D'Addabbo}, {D{\'e}sert},
  {Didelon}, {Doyle}, {Goupy}, {K{\"o}nyves}, {Kramer}, {Lagache}, {Leclercq},
  {Mac{\'\i}as-P{\'e}rez}, {Maury}, {Mauskopf}, {Mayet}, {Monfardini}, {Motte},
  {Pajot}, {Pascale}, {Peretto}, {Perotto}, {Pisano}, {Ponthieu},
  {Rev{\'e}ret}, {Rigby}, {Ritacco}, {Rodriguez}, {Romero}, {Roy}, {Ruppin},
  {Schuster}, {Sievers}, {Triqueneaux}, {Tucker}, \& {Zylka}}]{Bracco2017}
{Bracco}, A., {Palmeirim}, P., {Andr{\'e}}, P., {et~al.} 2017, \aap, 604, A52,
  \dodoi{10.1051/0004-6361/201731117}

\bibitem[{{Caselli} {et~al.}(1994){Caselli}, {Hasegawa}, \&
  {Herbst}}]{Caselli1994}
{Caselli}, P., {Hasegawa}, T.~I., \& {Herbst}, E. 1994, \apj, 421, 206,
  \dodoi{10.1086/173637}

\bibitem[{{Caselli} {et~al.}(1999){Caselli}, {Walmsley}, {Tafalla}, {Dore}, \&
  {Myers}}]{Caselli1999}
{Caselli}, P., {Walmsley}, C.~M., {Tafalla}, M., {Dore}, L., \& {Myers}, P.~C.
  1999, \apjl, 523, L165, \dodoi{10.1086/312280}

\bibitem[{{Caselli} {et~al.}(1998){Caselli}, {Walmsley}, {Terzieva}, \&
  {Herbst}}]{Caselli1998}
{Caselli}, P., {Walmsley}, C.~M., {Terzieva}, R., \& {Herbst}, E. 1998, \apj,
  499, 234, \dodoi{10.1086/305624}

\bibitem[{{Caselli} {et~al.}(2002){Caselli}, {Walmsley}, {Zucconi}, {Tafalla},
  {Dore}, \& {Myers}}]{Caselli2002-b}
{Caselli}, P., {Walmsley}, C.~M., {Zucconi}, A., {et~al.} 2002, \apj, 565, 344,
  \dodoi{10.1086/324302}

\bibitem[{{Chaabouni} {et~al.}(2012){Chaabouni}, {Bergeron}, {Baouche},
  {Dulieu}, {Matar}, {Congiu}, {Gavilan}, \& {Lemaire}}]{Chaabouni2012}
{Chaabouni}, H., {Bergeron}, H., {Baouche}, S., {et~al.} 2012, \aap, 538, A128,
  \dodoi{10.1051/0004-6361/201117409}

\bibitem[{{Chac{\'o}n-Tanarro} {et~al.}(2017){Chac{\'o}n-Tanarro}, {Caselli},
  {Bizzocchi}, {Pineda}, {Harju}, {Spaans}, \&
  {D{\'e}sert}}]{Chacon-Tanarro2017}
{Chac{\'o}n-Tanarro}, A., {Caselli}, P., {Bizzocchi}, L., {et~al.} 2017, \aap,
  606, A142, \dodoi{10.1051/0004-6361/201630265}

\bibitem[{{Chen} {et~al.}(2014){Chen}, {Chuang}, {Mu{\~n}oz Caro}, {Nuevo},
  {Chu}, {Yih}, {Ip}, \& {Wu}}]{Chen2014}
{Chen}, Y.~J., {Chuang}, K.~J., {Mu{\~n}oz Caro}, G.~M., {et~al.} 2014, \apj,
  781, 15, \dodoi{10.1088/0004-637X/781/1/15}

\bibitem[{{Chuang} {et~al.}(2018){Chuang}, {Fedoseev}, {Qasim}, {Ioppolo}, {van
  Dishoeck}, \& {Linnartz}}]{Chuang_ea18}
{Chuang}, K.~J., {Fedoseev}, G., {Qasim}, D., {et~al.} 2018, \apj, 853, 102,
  \dodoi{10.3847/1538-4357/aaa24e}

\bibitem[{{Crapsi} {et~al.}(2005){Crapsi}, {Caselli}, {Walmsley}, {Myers},
  {Tafalla}, {Lee}, \& {Bourke}}]{Crapsi2005}
{Crapsi}, A., {Caselli}, P., {Walmsley}, C.~M., {et~al.} 2005, \apj, 619, 379,
  \dodoi{10.1086/426472}

\bibitem[{{Cruz-Diaz} {et~al.}(2016){Cruz-Diaz}, {Mart{\'{\i}}n-Dom{\'e}nech},
  {Mu{\~n}oz Caro}, \& {Chen}}]{Cruz-Diaz2016}
{Cruz-Diaz}, G.~A., {Mart{\'{\i}}n-Dom{\'e}nech}, R., {Mu{\~n}oz Caro}, G.~M.,
  \& {Chen}, Y.-J. 2016, \aap, 592, A68, \dodoi{10.1051/0004-6361/201526761}

\bibitem[{{Draine} \& {Li}(2007)}]{Draine2007}
{Draine}, B.~T., \& {Li}, A. 2007, \apj, 657, 810, \dodoi{10.1086/511055}

\bibitem[{{Fayolle} {et~al.}(2011){Fayolle}, {Bertin}, {Romanzin}, {Michaut},
  {{\"O}berg}, {Linnartz}, \& {Fillion}}]{Fayolle2011}
{Fayolle}, E.~C., {Bertin}, M., {Romanzin}, C., {et~al.} 2011, \apjl, 739, L36,
  \dodoi{10.1088/2041-8205/739/2/L36}

\bibitem[{{Fedoseev} {et~al.}(2015){Fedoseev}, {Cuppen}, {Ioppolo}, {Lamberts},
  \& {Linnartz}}]{Fedoseev2015}
{Fedoseev}, G., {Cuppen}, H.~M., {Ioppolo}, S., {Lamberts}, T., \& {Linnartz},
  H. 2015, \mnras, 448, 1288, \dodoi{10.1093/mnras/stu2603}

\bibitem[{{Fraser} \& {van Dishoeck}(2004)}]{Fraser2004}
{Fraser}, H.~J., \& {van Dishoeck}, E.~F. 2004, Advances in Space Research, 33,
  14, \dodoi{10.1016/j.asr.2003.04.003}

\bibitem[{{Fredon} {et~al.}(2017){Fredon}, {Lamberts}, \&
  {Cuppen}}]{Fredon_ea17}
{Fredon}, A., {Lamberts}, T., \& {Cuppen}, H.~M. 2017, \apj, 849, 125,
  \dodoi{10.3847/1538-4357/aa8c05}

\bibitem[{{Frerking} {et~al.}(1982){Frerking}, {Langer}, \&
  {Wilson}}]{Frerking1982}
{Frerking}, M.~A., {Langer}, W.~D., \& {Wilson}, R.~W. 1982, \apj, 262, 590,
  \dodoi{10.1086/160451}

\bibitem[{{Friesen} {et~al.}(2017){Friesen}, {Pineda}, {co-PIs}, {Rosolowsky},
  {Alves}, {Chac{\'o}n-Tanarro}, {How-Huan Chen}, {Chun-Yuan Chen}, {Di
  Francesco}, {Keown}, {Kirk}, {Punanova}, {Seo}, {Shirley}, {Ginsburg},
  {Hall}, {Offner}, {Singh}, {Arce}, {Caselli}, {Goodman}, {Martin}, {Matzner},
  {Myers}, {Redaelli}, \& {The GAS Collaboration}}]{Friesen2017}
{Friesen}, R.~K., {Pineda}, J.~E., {co-PIs}, {et~al.} 2017, \apj, 843, 63,
  \dodoi{10.3847/1538-4357/aa6d58}

\bibitem[{{Fuchs} {et~al.}(2009){Fuchs}, {Cuppen}, {Ioppolo}, {Romanzin},
  {Bisschop}, {Andersson}, {van Dishoeck}, \& {Linnartz}}]{Fuchs_ea09}
{Fuchs}, G.~W., {Cuppen}, H.~M., {Ioppolo}, S., {et~al.} 2009, \aap, 505, 629,
  \dodoi{10.1051/0004-6361/200810784}

\bibitem[{{Garrod} {et~al.}(2006){Garrod}, {Park}, {Caselli}, \&
  {Herbst}}]{Garrod_ea06}
{Garrod}, R., {Park}, I.~H., {Caselli}, P., \& {Herbst}, E. 2006, Faraday
  Discussions, 133, 51, \dodoi{10.1039/b516202e}

\bibitem[{{Garrod} {et~al.}(2007){Garrod}, {Wakelam}, \& {Herbst}}]{Garrod2007}
{Garrod}, R.~T., {Wakelam}, V., \& {Herbst}, E. 2007, \aap, 467, 1103,
  \dodoi{10.1051/0004-6361:20066704}

\bibitem[{{Geppert} {et~al.}(2006){Geppert}, {Hamberg}, {Thomas},
  {{\"O}sterdahl}, {Hellberg}, {Zhaunerchyk}, {Ehlerding}, {Millar}, {Roberts},
  {Semaniak}, {Ugglas}, {K{\"a}llberg}, {Simonsson}, {Kaminska}, \&
  {Larsson}}]{Geppert_ea06}
{Geppert}, W.~D., {Hamberg}, M., {Thomas}, R.~D., {et~al.} 2006, Faraday
  Discussions, 133, 177, \dodoi{10.1039/B516010C}

\bibitem[{{Ginsburg} \& {Mirocha}(2011)}]{Ginsburg2011}
{Ginsburg}, A., \& {Mirocha}, J. 2011, {PySpecKit: Python Spectroscopic
  Toolkit}, Astrophysics Source Code Library.
\newblock \doeprint{1109.001}

\bibitem[{{Goldsmith}(2001)}]{Goldsmith2001}
{Goldsmith}, P.~F. 2001, \apj, 557, 736, \dodoi{10.1086/322255}

\bibitem[{{Goldsmith} \& {Langer}(1999)}]{Goldsmith1999}
{Goldsmith}, P.~F., \& {Langer}, W.~D. 1999, \apj, 517, 209,
  \dodoi{10.1086/307195}

\bibitem[{{Gordy} \& {Cook}(1970)}]{Gordy1970}
{Gordy}, W., \& {Cook}, R.~L. 1970, Chemical applications of spectroscopy,
  Vol.~2, {Microwave molecular spectra} (New York u.a.: Interscience Publisher,
  1970)

\bibitem[{{Graedel} {et~al.}(1982){Graedel}, {Langer}, \&
  {Frerking}}]{Graedel1982}
{Graedel}, T.~E., {Langer}, W.~D., \& {Frerking}, M.~A. 1982, \apjs, 48, 321,
  \dodoi{10.1086/190780}

\bibitem[{{G{\"u}ver} \& {{\"O}zel}(2009)}]{Guver2009}
{G{\"u}ver}, T., \& {{\"O}zel}, F. 2009, \mnras, 400, 2050,
  \dodoi{10.1111/j.1365-2966.2009.15598.x}

\bibitem[{{Hacar} {et~al.}(2013){Hacar}, {Tafalla}, {Kauffmann}, \&
  {Kov{\'a}cs}}]{Hacar2013}
{Hacar}, A., {Tafalla}, M., {Kauffmann}, J., \& {Kov{\'a}cs}, A. 2013, \aap,
  554, A55, \dodoi{10.1051/0004-6361/201220090}

\bibitem[{{Harju} {et~al.}(2017){Harju}, {Daniel}, {Sipil{\"a}}, {Caselli},
  {Pineda}, {Friesen}, {Punanova}, {G{\"u}sten}, {Wiesenfeld}, {Myers},
  {Faure}, {Hily-Blant}, {Rist}, {Rosolowsky}, {Schlemmer}, \&
  {Shirley}}]{Harju2017}
{Harju}, J., {Daniel}, F., {Sipil{\"a}}, O., {et~al.} 2017, \aap, 600, A61,
  \dodoi{10.1051/0004-6361/201628463}

\bibitem[{{Harju} {et~al.}(2020){Harju}, {Pineda}, {Vasyunin}, {Caselli},
  {Offner}, {Goodman}, {Juvela}, {Sipil{\"a}}, {Faure}, {Le Gal}, {Hily-Blant},
  {Alves}, {Bizzocchi}, {Burkert}, {Chen}, {Friesen}, {G{\"u}sten}, {Myers},
  {Punanova}, {Rist}, {Rosolowsky}, {Schlemmer}, {Shirley}, {Spezzano},
  {Vastel}, \& {Wiesenfeld}}]{Harju2020}
{Harju}, J., {Pineda}, J.~E., {Vasyunin}, A.~I., {et~al.} 2020, \apj, 895, 101,
  \dodoi{10.3847/1538-4357/ab8f93}

\bibitem[{{Hasegawa} \& {Herbst}(1993)}]{HasegawaHerbst93}
{Hasegawa}, T.~I., \& {Herbst}, E. 1993, \mnras, 261, 83,
  \dodoi{10.1093/mnras/261.1.83}

\bibitem[{{Hasegawa} {et~al.}(1992){Hasegawa}, {Herbst}, \&
  {Leung}}]{Hasegawa1992}
{Hasegawa}, T.~I., {Herbst}, E., \& {Leung}, C.~M. 1992, \apjs, 82, 167,
  \dodoi{10.1086/191713}

\bibitem[{{Hasenberger} \& {Alves}(2020)}]{HasenbergerAlves21}
{Hasenberger}, B., \& {Alves}, J. 2020, \aap, 633, A132,
  \dodoi{10.1051/0004-6361/201936095}

\bibitem[{{Jim{\'e}nez-Serra} {et~al.}(2021){Jim{\'e}nez-Serra}, {Vasyunin},
  {Spezzano}, {Caselli}, {Cosentino}, \& {Viti}}]{JimenezSerra_ea21}
{Jim{\'e}nez-Serra}, I., {Vasyunin}, A.~I., {Spezzano}, S., {et~al.} 2021,
  \apj, 917, 44, \dodoi{10.3847/1538-4357/ac024c}

\bibitem[{{Jim{\'e}nez-Serra} {et~al.}(2016){Jim{\'e}nez-Serra}, {Vasyunin},
  {Caselli}, {Marcelino}, {Billot}, {Viti}, {Testi}, {Vastel}, {Lefloch}, \&
  {Bachiller}}]{Jimenez-Serra2016}
{Jim{\'e}nez-Serra}, I., {Vasyunin}, A.~I., {Caselli}, P., {et~al.} 2016,
  \apjl, 830, L6, \dodoi{10.3847/2041-8205/830/1/L6}

\bibitem[{{Jin} \& {Garrod}(2020)}]{Jin2020}
{Jin}, M., \& {Garrod}, R.~T. 2020, \apjs, 249, 26,
  \dodoi{10.3847/1538-4365/ab9ec8}

\bibitem[{{J{\o}rgensen} {et~al.}(2002){J{\o}rgensen}, {Sch{\"o}ier}, \& {van
  Dishoeck}}]{Jorgensen2002}
{J{\o}rgensen}, J.~K., {Sch{\"o}ier}, F.~L., \& {van Dishoeck}, E.~F. 2002,
  \aap, 389, 908, \dodoi{10.1051/0004-6361:20020681}

\bibitem[{{Keto} \& {Caselli}(2008)}]{Keto2008}
{Keto}, E., \& {Caselli}, P. 2008, \apj, 683, 238, \dodoi{10.1086/589147}

\bibitem[{{Keto} \& {Caselli}(2010)}]{Keto2010}
---. 2010, \mnras, 402, 1625, \dodoi{10.1111/j.1365-2966.2009.16033.x}

\bibitem[{{Kirk} {et~al.}(2005){Kirk}, {Ward-Thompson}, \&
  {Andr{\'e}}}]{Kirk2005}
{Kirk}, J.~M., {Ward-Thompson}, D., \& {Andr{\'e}}, P. 2005, \mnras, 360, 1506,
  \dodoi{10.1111/j.1365-2966.2005.09145.x}

\bibitem[{{Laas} \& {Caselli}(2019)}]{LaasCaselli19}
{Laas}, J.~C., \& {Caselli}, P. 2019, \aap, 624, A108,
  \dodoi{10.1051/0004-6361/201834446}

\bibitem[{{Lacy} {et~al.}(1994){Lacy}, {Knacke}, {Geballe}, \&
  {Tokunaga}}]{Lacy1994}
{Lacy}, J.~H., {Knacke}, R., {Geballe}, T.~R., \& {Tokunaga}, A.~T. 1994,
  \apjl, 428, L69, \dodoi{10.1086/187395}

\bibitem[{{Lacy} {et~al.}(2017){Lacy}, {Sneden}, {Kim}, \& {Jaffe}}]{Lacy2017}
{Lacy}, J.~H., {Sneden}, C., {Kim}, H., \& {Jaffe}, D.~T. 2017, \apj, 838, 66,
  \dodoi{10.3847/1538-4357/aa6247}

\bibitem[{{Ladjelate} {et~al.}(2020){Ladjelate}, {Andr{\'e}}, {K{\"o}nyves},
  {Ward-Thompson}, {Men'shchikov}, {Bracco}, {Palmeirim}, {Roy}, {Shimajiri},
  {Kirk}, {Arzoumanian}, {Benedettini}, {Di Francesco}, {Fiorellino},
  {Schneider}, {Pezzuto}, {Motte}, \& {Herschel Gould Belt Survey
  Team}}]{Ladjelate2020}
{Ladjelate}, B., {Andr{\'e}}, P., {K{\"o}nyves}, V., {et~al.} 2020, \aap, 638,
  A74, \dodoi{10.1051/0004-6361/201936442}

\bibitem[{{Lamberts} \& {K{\"a}stner}(2017)}]{Lamberts_2017}
{Lamberts}, T., \& {K{\"a}stner}, J. 2017, \apj, 846, 43,
  \dodoi{10.3847/1538-4357/aa8311}

\bibitem[{{Lattanzi} {et~al.}(2020){Lattanzi}, {Bizzocchi}, {Vasyunin},
  {Harju}, {Giuliano}, {Vastel}, \& {Caselli}}]{Lattanzi2020}
{Lattanzi}, V., {Bizzocchi}, L., {Vasyunin}, A.~I., {et~al.} 2020, \aap, 633,
  A118, \dodoi{10.1051/0004-6361/201936884}

\bibitem[{{Lee} {et~al.}(1998){Lee}, {Roueff}, {Pineau des Forets},
  {Shalabiea}, {Terzieva}, \& {Herbst}}]{Lee1998}
{Lee}, H.~H., {Roueff}, E., {Pineau des Forets}, G., {et~al.} 1998, \aap, 334,
  1047

\bibitem[{{Lee} {et~al.}(2003){Lee}, {Evans}, {Shirley}, \&
  {Tatematsu}}]{Lee2003}
{Lee}, J.-E., {Evans}, Neal~J., I., {Shirley}, Y.~L., \& {Tatematsu}, K. 2003,
  \apj, 583, 789, \dodoi{10.1086/345428}

\bibitem[{{Lees} \& {Baker}(1968)}]{Lees1968}
{Lees}, R.~M., \& {Baker}, J.~G. 1968, The Journal of Chemical Physics, 48,
  5299, \dodoi{10.1063/1.1668221}

\bibitem[{{Loison} {et~al.}(2020){Loison}, {Wakelam}, {Gratier}, \&
  {Hickson}}]{Loison2020}
{Loison}, J.-C., {Wakelam}, V., {Gratier}, P., \& {Hickson}, K.~M. 2020,
  \mnras, 498, 4663, \dodoi{10.1093/mnras/staa2700}

\bibitem[{{Lynds}(1962)}]{Lynds1962}
{Lynds}, B.~T. 1962, \apjs, 7, 1, \dodoi{10.1086/190072}

\bibitem[{{Marsh} {et~al.}(2014){Marsh}, {Griffin}, {Palmeirim}, {Andr{\'e}},
  {Kirk}, {Stamatellos}, {Ward-Thompson}, {Roy}, {Bontemps}, {Francesco},
  {Elia}, {Hill}, {K{\"o}nyves}, {Motte}, {Nguyen-Luong}, {Peretto}, {Pezzuto},
  {Rivera-Ingraham}, {Schneider}, {Spinoglio}, \& {White}}]{Marsh2014}
{Marsh}, K.~A., {Griffin}, M.~J., {Palmeirim}, P., {et~al.} 2014, \mnras, 439,
  3683, \dodoi{10.1093/mnras/stu219}

\bibitem[{{Minissale} {et~al.}(2016){Minissale}, {Moudens}, {Baouche},
  {Chaabouni}, \& {Dulieu}}]{Minissale2016}
{Minissale}, M., {Moudens}, A., {Baouche}, S., {Chaabouni}, H., \& {Dulieu}, F.
  2016, \mnras, 458, 2953, \dodoi{10.1093/mnras/stw373}

\bibitem[{{Mu{\~n}oz Caro} {et~al.}(2016){Mu{\~n}oz Caro}, {Chen}, {Aparicio},
  {Jim{\'e}nez-Escobar}, {Rosu-Finsen}, {Lasne}, \&
  {McCoustra}}]{MunozCaro2016}
{Mu{\~n}oz Caro}, G.~M., {Chen}, Y.~J., {Aparicio}, S., {et~al.} 2016, \aap,
  589, A19, \dodoi{10.1051/0004-6361/201628121}

\bibitem[{{Nagy} {et~al.}(2019){Nagy}, {Spezzano}, {Caselli}, {Vasyunin},
  {Tafalla}, {Bizzocchi}, {Prudenzano}, \& {Redaelli}}]{Nagy2019}
{Nagy}, Z., {Spezzano}, S., {Caselli}, P., {et~al.} 2019, \aap, 630, A136,
  \dodoi{10.1051/0004-6361/201935568}

\bibitem[{{{\"O}berg} {et~al.}(2009){{\"O}berg}, {Garrod}, {van Dishoeck}, \&
  {Linnartz}}]{Oeberg2009}
{{\"O}berg}, K.~I., {Garrod}, R.~T., {van Dishoeck}, E.~F., \& {Linnartz}, H.
  2009, \aap, 504, 891, \dodoi{10.1051/0004-6361/200912559}

\bibitem[{{Paardekooper} {et~al.}(2016){Paardekooper}, {Fedoseev}, {Riedo}, \&
  {Linnartz}}]{Paardekooper2016}
{Paardekooper}, D.~M., {Fedoseev}, G., {Riedo}, A., \& {Linnartz}, H. 2016,
  \aap, 596, A72, \dodoi{10.1051/0004-6361/201629063}

\bibitem[{{Pagani} {et~al.}(2007){Pagani}, {Bacmann}, {Cabrit}, \&
  {Vastel}}]{Pagani2007}
{Pagani}, L., {Bacmann}, A., {Cabrit}, S., \& {Vastel}, C. 2007, \aap, 467,
  179, \dodoi{10.1051/0004-6361:20066670}

\bibitem[{{Palmeirim} {et~al.}(2013){Palmeirim}, {Andr{\'e}}, {Kirk},
  {Ward-Thompson}, {Arzoumanian}, {K{\"o}nyves}, {Didelon}, {Schneider},
  {Benedettini}, {Bontemps}, {Di Francesco}, {Elia}, {Griffin}, {Hennemann},
  {Hill}, {Martin}, {Men'shchikov}, {Molinari}, {Motte}, {Nguyen Luong},
  {Nutter}, {Peretto}, {Pezzuto}, {Roy}, {Rygl}, {Spinoglio}, \&
  {White}}]{Palmeirim2013}
{Palmeirim}, P., {Andr{\'e}}, P., {Kirk}, J., {et~al.} 2013, \aap, 550, A38,
  \dodoi{10.1051/0004-6361/201220500}

\bibitem[{{Pickett} {et~al.}(1998){Pickett}, {Poynter}, {Cohen}, {Delitsky},
  {Pearson}, \& {M{\"u}ller}}]{Pickett1998}
{Pickett}, H.~M., {Poynter}, R.~L., {Cohen}, E.~A., {et~al.} 1998, \jqsrt, 60,
  883, \dodoi{10.1016/S0022-4073(98)00091-0}

\bibitem[{{Punanova} {et~al.}(2018{\natexlab{a}}){Punanova}, {Caselli},
  {Pineda}, {Pon}, {Tafalla}, {Hacar}, \& {Bizzocchi}}]{Punanova2018b}
{Punanova}, A., {Caselli}, P., {Pineda}, J.~E., {et~al.} 2018{\natexlab{a}},
  \aap, 617, A27, \dodoi{10.1051/0004-6361/201731159}

\bibitem[{{Punanova} {et~al.}(2018{\natexlab{b}}){Punanova}, {Caselli}, {Feng},
  {Chac{\'o}n-Tanarro}, {Ceccarelli}, {Neri}, {Fontani}, {Jim{\'e}nez-Serra},
  {Vastel}, {Bizzocchi}, {Pon}, {Vasyunin}, {Spezzano}, {Hily-Blant}, {Testi},
  {Viti}, {Yamamoto}, {Alves}, {Bachiller}, {Balucani}, {Bianchi},
  {Bottinelli}, {Caux}, {Choudhury}, {Codella}, {Dulieu}, {Favre}, {Holdship},
  {Jaber Al-Edhari}, {Kahane}, {Laas}, {LeFloch}, {L{\'o}pez-Sepulcre},
  {Ospina-Zamudio}, {Oya}, {Pineda}, {Podio}, {Quenard}, {Rimola}, {Sakai},
  {Sims}, {Taquet}, {Theul{\'e}}, \& {Ugliengo}}]{Punanova2018a}
{Punanova}, A., {Caselli}, P., {Feng}, S., {et~al.} 2018{\natexlab{b}}, \apj,
  855, 112, \dodoi{10.3847/1538-4357/aaad09}

\bibitem[{{Rebull} {et~al.}(2010){Rebull}, {Padgett}, {McCabe}, {Hillenbrand},
  {Stapelfeldt}, {Noriega-Crespo}, {Carey}, {Brooke}, {Huard}, {Terebey},
  {Audard}, {Monin}, {Fukagawa}, {G{\"u}del}, {Knapp}, {Menard}, {Allen},
  {Angione}, {Baldovin-Saavedra}, {Bouvier}, {Briggs}, {Dougados}, {Evans},
  {Flagey}, {Guieu}, {Grosso}, {Glauser}, {Harvey}, {Hines}, {Latter},
  {Skinner}, {Strom}, {Tromp}, \& {Wolf}}]{Rebull2010}
{Rebull}, L.~M., {Padgett}, D.~L., {McCabe}, C.-E., {et~al.} 2010, \apjs, 186,
  259, \dodoi{10.1088/0067-0049/186/2/259}

\bibitem[{{Rimola} {et~al.}(2014){Rimola}, {Taquet}, {Ugliengo}, {Balucani}, \&
  {Ceccarelli}}]{Rimola2014}
{Rimola}, A., {Taquet}, V., {Ugliengo}, P., {Balucani}, N., \& {Ceccarelli}, C.
  2014, \aap, 572, A70, \dodoi{10.1051/0004-6361/201424046}

\bibitem[{{Roccatagliata} {et~al.}(2020){Roccatagliata}, {Franciosini},
  {Sacco}, {Rand ich}, \& {Sicilia-Aguilar}}]{Roccatagliata2020}
{Roccatagliata}, V., {Franciosini}, E., {Sacco}, G.~G., {Rand ich}, S., \&
  {Sicilia-Aguilar}, A. 2020, \aap, 638, A85,
  \dodoi{10.1051/0004-6361/201936401}

\bibitem[{{Ruaud} {et~al.}(2016){Ruaud}, {Wakelam}, \& {Hersant}}]{Ruaud2016}
{Ruaud}, M., {Wakelam}, V., \& {Hersant}, F. 2016, \mnras, 459, 3756,
  \dodoi{10.1093/mnras/stw887}

\bibitem[{{Santiago-Garc{\'{\i}}a} {et~al.}(2009){Santiago-Garc{\'{\i}}a},
  {Tafalla}, {Johnstone}, \& {Bachiller}}]{Santiago-Garcia2009}
{Santiago-Garc{\'{\i}}a}, J., {Tafalla}, M., {Johnstone}, D., \& {Bachiller},
  R. 2009, \aap, 495, 169, \dodoi{10.1051/0004-6361:200810739}

\bibitem[{{Schlafly} {et~al.}(2014){Schlafly}, {Green}, {Finkbeiner}, {Rix},
  {Bell}, {Burgett}, {Chambers}, {Draper}, {Hodapp}, {Kaiser}, {Magnier},
  {Martin}, {Metcalfe}, {Price}, \& {Tonry}}]{Schlafly2014}
{Schlafly}, E.~F., {Green}, G., {Finkbeiner}, D.~P., {et~al.} 2014, \apj, 786,
  29, \dodoi{10.1088/0004-637X/786/1/29}

\bibitem[{{Schmalzl} {et~al.}(2010){Schmalzl}, {Kainulainen}, {Quanz}, {Alves},
  {Goodman}, {Henning}, {Launhardt}, {Pineda}, \&
  {Rom{\'a}n-Z{\'u}{\~n}iga}}]{Schmalzl2010}
{Schmalzl}, M., {Kainulainen}, J., {Quanz}, S.~P., {et~al.} 2010, \apj, 725,
  1327, \dodoi{10.1088/0004-637X/725/1/1327}

\bibitem[{{Schnee} {et~al.}(2010){Schnee}, {Enoch}, {Noriega-Crespo}, {Sayers},
  {Terebey}, {Caselli}, {Foster}, {Goodman}, {Kauffmann}, {Padgett}, {Rebull},
  {Sargent}, \& {Shetty}}]{Schnee2010}
{Schnee}, S., {Enoch}, M., {Noriega-Crespo}, A., {et~al.} 2010, \apj, 708, 127,
  \dodoi{10.1088/0004-637X/708/1/127}

\bibitem[{{Sch{\"o}ier} {et~al.}(2005){Sch{\"o}ier}, {van der Tak}, {van
  Dishoeck}, \& {Black}}]{Schoeier2005}
{Sch{\"o}ier}, F.~L., {van der Tak}, F.~F.~S., {van Dishoeck}, E.~F., \&
  {Black}, J.~H. 2005, \aap, 432, 369, \dodoi{10.1051/0004-6361:20041729}

\bibitem[{{Scibelli} \& {Shirley}(2020)}]{Scibelli2020}
{Scibelli}, S., \& {Shirley}, Y. 2020, \apj, 891, 73,
  \dodoi{10.3847/1538-4357/ab7375}

\bibitem[{{Scibelli} {et~al.}(2021){Scibelli}, {Shirley}, {Vasyunin}, \&
  {Launhardt}}]{Scibelli2021}
{Scibelli}, S., {Shirley}, Y., {Vasyunin}, A., \& {Launhardt}, R. 2021, \mnras,
  504, 5754, \dodoi{10.1093/mnras/stab1151}

\bibitem[{{Seo} {et~al.}(2015){Seo}, {Shirley}, {Goldsmith}, {Ward-Thompson},
  {Kirk}, {Schmalzl}, {Lee}, {Friesen}, {Langston}, {Masters}, \&
  {Garwood}}]{Seo2015}
{Seo}, Y.~M., {Shirley}, Y.~L., {Goldsmith}, P., {et~al.} 2015, \apj, 805, 185,
  \dodoi{10.1088/0004-637X/805/2/185}

\bibitem[{{Seo} {et~al.}(2019){Seo}, {Majumdar}, {Goldsmith}, {Shirley},
  {Willacy}, {Ward-Thompson}, {Friesen}, {Frayer}, {Church}, {Chung}, {Cleary},
  {Cunningham}, {Devaraj}, {Egan}, {Gaier}, {Gawande}, {Gundersen}, {Harris},
  {Kangaslahti}, {Readhead}, {Samoska}, {Sieth}, {Stennes}, {Voll}, \&
  {White}}]{Seo2019}
{Seo}, Y.~M., {Majumdar}, L., {Goldsmith}, P.~F., {et~al.} 2019, \apj, 871,
  134, \dodoi{10.3847/1538-4357/aaf887}

\bibitem[{{Shalabiea}(2001)}]{Shalabiea2001}
{Shalabiea}, O.~M. 2001, \aap, 370, 1044, \dodoi{10.1051/0004-6361:20010323}

\bibitem[{{Shingledecker} {et~al.}(2020){Shingledecker}, {Lamberts}, {Laas},
  {Vasyunin}, {Herbst}, {K{\"a}stner}, \& {Caselli}}]{Shingledecker_ea20}
{Shingledecker}, C.~N., {Lamberts}, T., {Laas}, J.~C., {et~al.} 2020, \apj,
  888, 52, \dodoi{10.3847/1538-4357/ab5360}

\bibitem[{{Sipil{\"a}} {et~al.}(2020){Sipil{\"a}}, {Zhao}, \&
  {Caselli}}]{Sipila2020}
{Sipil{\"a}}, O., {Zhao}, B., \& {Caselli}, P. 2020, \aap, 640, A94,
  \dodoi{10.1051/0004-6361/202038353}

\bibitem[{{Spezzano} {et~al.}(2016){Spezzano}, {Bizzocchi}, {Caselli}, {Harju},
  \& {Br{\"u}nken}}]{Spezzano2016l}
{Spezzano}, S., {Bizzocchi}, L., {Caselli}, P., {Harju}, J., \& {Br{\"u}nken},
  S. 2016, \aap, 592, L11, \dodoi{10.1051/0004-6361/201628652}

\bibitem[{{Spezzano} {et~al.}(2020){Spezzano}, {Caselli}, {Pineda},
  {Bizzocchi}, {Prudenzano}, \& {Nagy}}]{Spezzano_ea20}
{Spezzano}, S., {Caselli}, P., {Pineda}, J.~E., {et~al.} 2020, \aap, 643, A60,
  \dodoi{10.1051/0004-6361/201936598}

\bibitem[{{Tafalla} \& {Hacar}(2015)}]{Tafalla2015}
{Tafalla}, M., \& {Hacar}, A. 2015, \aap, 574, A104,
  \dodoi{10.1051/0004-6361/201424576}

\bibitem[{{Tafalla} \& {Santiago}(2004)}]{Tafalla2004b}
{Tafalla}, M., \& {Santiago}, J. 2004, \aap, 414, L53,
  \dodoi{10.1051/0004-6361:20031766}

\bibitem[{{Tafalla} {et~al.}(2006){Tafalla}, {Santiago-Garc{\'{\i}}a}, {Myers},
  {Caselli}, {Walmsley}, \& {Crapsi}}]{Tafalla2006}
{Tafalla}, M., {Santiago-Garc{\'{\i}}a}, J., {Myers}, P.~C., {et~al.} 2006,
  \aap, 455, 577, \dodoi{10.1051/0004-6361:20065311}

\bibitem[{{Vastel} {et~al.}(2014){Vastel}, {Ceccarelli}, {Lefloch}, \&
  {Bachiller}}]{Vastel2014}
{Vastel}, C., {Ceccarelli}, C., {Lefloch}, B., \& {Bachiller}, R. 2014, \apjl,
  795, L2, \dodoi{10.1088/2041-8205/795/1/L2}

\bibitem[{{Vasyunin} {et~al.}(2017){Vasyunin}, {Caselli}, {Dulieu}, \&
  {Jim{\'e}nez-Serra}}]{Vasyunin2017}
{Vasyunin}, A.~I., {Caselli}, P., {Dulieu}, F., \& {Jim{\'e}nez-Serra}, I.
  2017, \apj, 842, 33, \dodoi{10.3847/1538-4357/aa72ec}

\bibitem[{{Vasyunin} \& {Herbst}(2013{\natexlab{a}})}]{Vasyunin2013}
{Vasyunin}, A.~I., \& {Herbst}, E. 2013{\natexlab{a}}, \apj, 769, 34,
  \dodoi{10.1088/0004-637X/769/1/34}

\bibitem[{{Vasyunin} \& {Herbst}(2013{\natexlab{b}})}]{VasyuninHerbst13mc}
---. 2013{\natexlab{b}}, \apj, 762, 86, \dodoi{10.1088/0004-637X/762/2/86}

\bibitem[{{Wakelam} \& {Herbst}(2008)}]{Wakelam2008}
{Wakelam}, V., \& {Herbst}, E. 2008, \apj, 680, 371, \dodoi{10.1086/587734}

\bibitem[{{Wakelam} {et~al.}(2017){Wakelam}, {Loison}, {Mereau}, \&
  {Ruaud}}]{Wakelam_ea17}
{Wakelam}, V., {Loison}, J.~C., {Mereau}, R., \& {Ruaud}, M. 2017, Molecular
  Astrophysics, 6, 22, \dodoi{10.1016/j.molap.2017.01.002}

\bibitem[{{Walsh} {et~al.}(2009){Walsh}, {Harada}, {Herbst}, \&
  {Millar}}]{Walsh2009}
{Walsh}, C., {Harada}, N., {Herbst}, E., \& {Millar}, T.~J. 2009, \apj, 700,
  752, \dodoi{10.1088/0004-637X/700/1/752}

\bibitem[{{Wannier}(1980)}]{Wannier1980}
{Wannier}, P.~G. 1980, \araa, 18, 399,
  \dodoi{10.1146/annurev.aa.18.090180.002151}

\bibitem[{{Ward-Thompson} {et~al.}(2016){Ward-Thompson}, {Pattle}, {Kirk},
  {Marsh}, {Buckle}, {Hatchell}, {Nutter}, {Griffin}, {Di Francesco},
  {Andr{\'e}}, {Beaulieu}, {Berry}, {Broekhoven-Fiene}, {Currie}, {Fich},
  {Jenness}, {Johnstone}, {Kirk}, {Mottram}, {Pineda}, {Quinn}, {Sadavoy},
  {Salji}, {Tisi}, {Walker-Smith}, {White}, {Hill}, {K{\"o}nyves}, {Palmeirim},
  \& {Pezzuto}}]{Ward-Thompson2016}
{Ward-Thompson}, D., {Pattle}, K., {Kirk}, J.~M., {et~al.} 2016, \mnras, 463,
  1008, \dodoi{10.1093/mnras/stw1978}

\bibitem[{{Watanabe} \& {Kouchi}(2002)}]{Watanabe2002}
{Watanabe}, N., \& {Kouchi}, A. 2002, \apjl, 571, L173, \dodoi{10.1086/341412}

\bibitem[{{Whittet} {et~al.}(2011){Whittet}, {Cook}, {Herbst}, {Chiar}, \&
  {Shenoy}}]{Whittet_2011}
{Whittet}, D.~C.~B., {Cook}, A.~M., {Herbst}, E., {Chiar}, J.~E., \& {Shenoy},
  S.~S. 2011, \apj, 742, 28, \dodoi{10.1088/0004-637X/742/1/28}

\bibitem[{{Willacy} {et~al.}(1998){Willacy}, {Langer}, \&
  {Velusamy}}]{Willacy_ea98}
{Willacy}, K., {Langer}, W.~D., \& {Velusamy}, T. 1998, \apjl, 507, L171,
  \dodoi{10.1086/311695}

\bibitem[{{Wilson} \& {Rood}(1994)}]{Wilson1994}
{Wilson}, T.~L., \& {Rood}, R. 1994, \araa, 32, 191,
  \dodoi{10.1146/annurev.aa.32.090194.001203}

\bibitem[{{Wirstr{\"o}m} {et~al.}(2011){Wirstr{\"o}m}, {Geppert}, {Hjalmarson},
  {Persson}, {Black}, {Bergman}, {Millar}, {Hamberg}, \&
  {Vigren}}]{Wirstrom2011}
{Wirstr{\"o}m}, E.~S., {Geppert}, W.~D., {Hjalmarson}, {\r{A}}., {et~al.} 2011,
  \aap, 533, A24, \dodoi{10.1051/0004-6361/201116525}

\bibitem[{{Xu} \& {Lovas}(1997)}]{Xu1997}
{Xu}, L.-H., \& {Lovas}, F.~J. 1997, Journal of Physical and Chemical Reference
  Data, 26, 17, \dodoi{10.1063/1.556005}

\end{thebibliography}
\bibliographystyle{aasjournal}



\end{document}